\documentclass[twocolumn]{aastex63}

\usepackage{natbib}
\usepackage[style=american]{csquotes}
\usepackage{lipsum}  

\usepackage[hyphens]{url}

\graphicspath{{./}{Figures/}}

\begin{document}

\title{Giant Outer Transiting Exoplanet Mass (GOT 'EM) Survey: III. Recovery and Confirmation of a Temperate, Mildly Eccentric, Single-Transit Jupiter Orbiting TOI-2010.}

\shorttitle{GOT 'EM: III} % shows at top of every page
\shortauthors{Mann et al.}

\correspondingauthor{Christopher Mann}
\email{christopher.mann@umontreal.ca}

%% Author list (quite long so in a different file)

         %% ---------------------- %%
%% ----- %% ---- CORE AUTHORS ---- %% ----- %%
         %% ---------------------- %%

\author[0000-0002-9312-0073]{Christopher R. Mann}
\affiliation{Département de Physique, Université de Montréal, Montréal, QC, Canada}
\affiliation{Trottier Institute for Research on Exoplanets (\emph{iREx})}

\author[0000-0002-4297-5506]{Paul A.\ Dalba}
\altaffiliation{Heising-Simons 51 Pegasi b Postdoctoral Fellow.}
\affiliation{Department of Astronomy and Astrophysics, University of California, Santa Cruz, CA 95064, USA}
\affiliation{SETI Institute, Carl Sagan Center, 339 Bernardo Ave, Suite 200, Mountain View, CA 94043, USA}

\author[0000-0002-6780-4252]{David Lafrenière}
\affiliation{Département de Physique, Université de Montréal, Montréal, QC, Canada}
\affiliation{Trottier Institute for Research on Exoplanets (\emph{iREx})}

         %% ------------------------ %%
%% ----- %% ---- APF PROCESSING ---- %% ----- %%
         %% ------------------------ %%

\author[0000-0003-3504-5316]{Benjamin J.\ Fulton}% (bjfulton@ipac.caltech.edu)
\affiliation{Cahill Center for Astronomy $\&$ Astrophysics, California Institute of Technology, Pasadena, CA 91125, USA}
\affiliation{IPAC-NASA Exoplanet Science Institute, Pasadena, CA 91125, USA}

         %% ---------------------- %%
%% ----- %% ---- SOPHIE CREW  ---- %% ----- %%
         %% ---------------------- %%
\author{Guillaume H\'ebrard}% <hebrard@iap.fr>
\affiliation{Institut d'astrophysique de Paris, UMR7095 CNRS, Universit\'e Pierre \& Marie Curie, 98bis boulevard Arago, 75014 Paris, France}
\affiliation{Observatoire de Haute-Provence, CNRS, Universit\'e d'Aix-Marseille, 04870 Saint-Michel-l'Observatoire, France}

\author{Isabelle Boisse}% <isabelle.boisse@lam.fr>
\affiliation{Laboratoire d'Astrophysique de Marseille, Universit\'e de Provence, UMR6110 CNRS, 38 rue F. Joliot Curie, 13388 Marseille cedex 13, France}

\author{Shweta Dalal}% <shweta.dalal@iap.fr>
\affiliation{Institut d'astrophysique de Paris, UMR7095 CNRS, Universit\'e Pierre \& Marie Curie, 98bis boulevard Arago, 75014 Paris, France}
\affiliation{Department of Astrophysics, University of Exeter, Stocker Rd, Exeter, EX4 4QL, UK}

\author{Magali Deleuil}% <magali.deleuil@lam.fr>
\affiliation{Laboratoire d'Astrophysique de Marseille, Universit\'e de Provence, UMR6110 CNRS, 38 rue F. Joliot Curie, 13388 Marseille cedex 13, France}

\author[0000-0001-5099-7978]{Xavier Delfosse}% <xavier.delfosse@univ-grenoble-alpes.fr>
\affiliation{Univ. Grenoble Alpes, CNRS, IPAG, 38000 Grenoble, France}

\author{Olivier Demangeon}%<olivier.demangeon@astro.up.pt>
\affiliation{Instituto de Astrof{\'\i}sica e Ci\^encias do Espa\c{c}o, Universidade do Porto, CAUP, Rua das Estrelas, 4150-762 Porto, Portugal}

\author[0000-0003-0536-4607]{Thierry Forveille}% <thierry.forveille@univ-grenoble-alpes.fr>
\affiliation{Universit\'e Grenoble Alpes, CNRS, IPAG, 38000 Grenoble, France}

\author{Neda Heidari}% <neda.heidari@lam.fr>
\affiliation{Laboratoire d'Astrophysique de Marseille, Universit\'e de Provence, UMR6110 CNRS, 38 rue F. Joliot Curie, 13388 Marseille cedex 13, France}

\author{Flavien Kiefer}% <kiefer@iap.fr>
\affiliation{LESIA, Observatoire de Paris, Universit\'e PSL, CNRS, Sorbonne Universit\'e, Universit\'e de Paris, 5 place Jules Janssen, 92195 Meudon, France}

\author[0000-0002-5084-168X]{Eder Martioli}% <martioli@iap.fr>
\affiliation{Laborat\'{o}rio Nacional de Astrof\'{i}sica, Rua Estados Unidos 154, 37504-364, Itajub\'{a} - MG, Brazil}
\affiliation{Institut d'astrophysique de Paris, UMR7095 CNRS, Universit\'e Pierre \& Marie Curie, 98bis boulevard Arago, 75014 Paris, France}

\author[0000-0002-2842-3924]{Claire Moutou}% <claire.moutou@irap.omp.eu>
\affiliation{Universit\'e de Toulouse, CNRS, IRAP, 14 avenue Belin, 31400 Toulouse, France}

         %% ---------------------- %%
%% ----- %% ---- TULL CREW  ---- %% ----- %%
         %% ---------------------- %%
\author[0000-0002-7714-6310]{Michael Endl}% mike@astro.as.utexas.edu
\affiliation{McDonald Observatory and Center for Planetary Systems Habitability, The University of Texas at Austin, Austin, TX 78730, USA}

\author[0000-0001-9662-3496]{William D. Cochran}% wdc@astro.as.utexas.edu
\affiliation{McDonald Observatory and Center for Planetary Systems Habitability, 
The University of Texas, Austin Texas, USA}

\author{Phillip MacQueen}% pjm@astro.as.utexas.edu
\affiliation{McDonald Observatory, The University of Texas at Austin, 2515 Speedway Blvd., Stop C1400, Austin Texas 78712, USA}
\author[0000-0001-7016-7277]{Franck Marchis}% fmarchis@seti.org
\affiliation{SETI Institute, Carl Sagan Center, 339 Bernardo Ave, Suite 200, Mountain View, CA 94043, USA}
\affiliation{Laboratoire d'Astrophysique de Marseille, Universit\'e de Provence, UMR6110 CNRS, 38 rue F. Joliot Curie, 13388 Marseille cedex 13, France}

         %% ---------------------------- %%
%% ----- %% ---- TSTPC CONTRIBUTERS ---- %% ----- %%
         %% ---------------------------- %%
\author[0000-0003-2313-467X]{Diana Dragomir}% - dragomir@unm.edu
\affiliation{Department of Physics and Astronomy, University of New Mexico, 1919 Lomas Blvd NE Albuquerque, NM, 87131, USA}

\author[0000-0002-5463-9980]{Arvind F.\ Gupta}% - arvind@psu.edu
\affil{Department of Astronomy \& Astrophysics, 525 Davey Laboratory, The Pennsylvania State University, University Park, PA, 16802, USA}
\affil{Center for Exoplanets and Habitable Worlds, 525 Davey Laboratory, The Pennsylvania State University, University Park, PA, 16802, USA}

\author[0000-0002-2457-7889]{Dax L. Feliz}% - dfeliz@amnh.org
\affiliation{American Museum of Natural History, 200 Central Park West, Manhattan, NY 10024, USA}

\author[0000-0003-1360-4404]{Belinda A. Nicholson}% - Belinda.Nicholson@usq.edu.au 
\affiliation{Centre for Astrophysics, University of Southern Queensland, Toowoomba, Australia, 4350}
\affiliation{Sub-department of Astrophysics, University of Oxford, Keble Rd, Oxford, United Kingdom, OX13RH}

\author{Carl Ziegler}% - Carl.Ziegler@sfasu.edu
\affiliation{Department of Physics, Engineering and Astronomy, Stephen F. Austin State University, 1936 North St, Nacogdoches, TX 75962, USA}

\author[0000-0001-6213-8804]{Steven Villanueva Jr.}% - steven.villanueva@nasa.gov
\altaffiliation{NPP Fellow.}
\affiliation{NASA Goddard Space Flight Center, Exoplanets and Stellar Astrophysics Laboratory (Code 667), Greenbelt, MD 20771, USA}

         %% ------------------------- %%
%% ----- %% ---- NEOSSat PEOPLE ---- %% ----- %%
         %% ------------------------- %%
\author[0000-0002-5904-1865]{Jason Rowe}% - jrowe@ubishops.ca
\affiliation{Bishops University, 2600 College St, Sherbrooke, QC J1M 1Z7, Canada}
\author[0000-0003-4787-2335]{Geert Jan Talens}% - gt8538@princeton.edu 
\affiliation{Department of Astrophysical Sciences, Princeton University, 4 Ivy Lane, Princeton, NJ 08544, USA}
\affiliation{Trottier Institute for Research on Exoplanets (\emph{iREx})}

         %% ------------------------- %%
%% ----- %% ---- METALICITY WORK ---- %% ----- %%
         %% ------------------------- %%
\author[0000-0002-5113-8558]{Daniel Thorngren}% daniel.thorngren@umontreal.ca
\affiliation{Department of Physics \& Astronomy, Johns Hopkins University, Baltimore, MD, 21210 USA}

         %% -------------------- %%
%% ----- %% ---- DISCOVERY ---- %% ----- %%
         %% -------------------- %%
\author[0000-0002-8527-2114]{Daryll LaCourse}% (nhawkb@gmail.com)
\affiliation{Amateur Astronomer, 7507 52nd Pl NE, Marysville, WA 98270, USA}
\author[0000-0003-3988-3245]{Tom Jacobs}% (tomjacobs128@gmail.com) 
\affiliation{Amateur Astronomer, 12812 SE 69th Place, Bellevue WA 98006, USA}

         %% ------------------------ %%
%% ----- %% ---- HIRES TEMPLATE ---- %% ----- %%
         %% ------------------------ %%
\author[0000-0001-8638-0320]{Andrew W.\ Howard}% ahoward@caltech.edu
\affiliation{Department of Astronomy, California Institute of Technology, Pasadena, CA 91125, USA}

         %% ----------------------- %%
%% ----- %% ---- TRES VETTING ---- %% ----- %%
         %% ----------------------- %%

\author[0000-0001-6637-5401]{Allyson~Bieryla}% abieryla@cfa.harvard.edu (TRES)
\affiliation{Center for Astrophysics ${\rm \mid}$ Harvard {\rm \&} Smithsonian, 60 Garden Street, Cambridge, MA 02138, USA}

\author[0000-0001-9911-7388]{David~W.~Latham}% - dlatham@cfa.harvard.edu (TRES) + TESS Architect 
\affiliation{Center for Astrophysics ${\rm \mid}$ Harvard {\rm \&} Smithsonian, 60 Garden Street, Cambridge, MA 02138, USA}

         %% --------------------------- %%
%% ----- %% ---- LCO/NRES SPECTRUM ---- %% ----- %%
         %% --------------------------- %%
\author[0000-0003-2935-7196]{Markus Rabus}% - mrabus@ucsc.cl 
\affiliation{Departamento de Matemática y Física Aplicadas, Facultad de Ingeniería, Universidad Católica de la Santísima Concepción, Alonso de Rivera 2850, Concepción, Chile}

         %% -------------------------------- %%
%% ----- %% ---- LIGHT CURVE MODULATION ---- %% ----- %%
         %% -------------------------------- %%
%% TESS
\author[0000-0002-3551-279X]{Tara Fetherolf}% tara.fetherolf@gmail.com
\altaffiliation{UC Chancellor's Fellow.}
\affiliation{Department of Earth and Planetary Sciences, University of California Riverside, 900 University Avenue, Riverside, CA 92521, USA}

%% WASP
\author{Coel Hellier}% c.hellier@keele.ac.uk (WASP)
\affiliation{Astrophysics Group, Keele University, Staffordshire ST5 5BG, U.K.}

         %% --------------------------- %%
%% ----- %% ---- 'ALOPEKE SPECKLE ---- %% ----- %%
         %% --------------------------- %%
\author[0000-0002-2532-2853]{Steve~B.~Howell}%  - steve.b.howell@nasa.gov
\affiliation{NASA Ames Research Center, Moffett Field, CA 94035, USA}

         %% ---------------------- %%
%% ----- %% ----   GMU CREW   ---- %% ----- %%
         %% ---------------------- %%
\author[0000-0002-8864-1667]{Peter Plavchan}% - plavchan@gmail.com
\affiliation{Department of Physics \& Astronomy, George Mason University, 4400 University Drive MS 3F3, Fairfax, VA 22030, USA}

\author{Michael Reefe}% - mreefe@mit.edu
\affiliation{Kavli Institute for Astrophysics and Space Research, Massachusetts Institute of Technology, 77 Massachusetts Avenue, Cambridge, MA 02139, USA}
\affiliation{Department of Physics \& Astronomy, George Mason University, 4400 University Drive MS 3F3, Fairfax, VA 22030, USA}

\author{Deven Combs}% - dcombs4@gmu.edu  
\affiliation{Department of Physics \& Astronomy, George Mason University, 4400 University Drive MS 3F3, Fairfax, VA 22030, USA}

\author{Michael Bowen}% - mbowen11@gmu.edu
\affiliation{Department of Physics \& Astronomy, George Mason University, 4400 University Drive MS 3F3, Fairfax, VA 22030, USA}

\author[0000-0002-7424-9891]{Justin Wittrock}% - jwittroc@gmu.edu
\affiliation{Department of Physics \& Astronomy, George Mason University, 4400 University Drive MS 3F3, Fairfax, VA 22030, USA}

%          %% ------------------------- %%
% %% ----- %% ---- SHANE/SHARKS AO ---- %% ----- %%
%          %% ------------------------- %%
% \author{Courtney Dressing}% - dressing@berkeley.edu
% \affiliation{\cmc{Unknown Affiliation}}
% \author{Steven Giacalone}% - steven_giacalone@berkeley.edu
% \affiliation{\cmc{Unknown Affiliation}}

         %% ------------------------- %%
%% ----- %% ---- TESS ARCHITECTS ---- %% ----- %%
         %% ------------------------- %%

%% David Latham, included above for other contributions, would normally be on this list

\author[0000-0003-2058-6662]{George~R.~Ricker} % TESS Architect
\affiliation{Department of Physics and Kavli Institute for Astrophysics and Space Research, Massachusetts Institute of Technology, Cambridge, MA 02139, USA}

% \author[0000-0001-6763-6562]{Roland~Vanderspek} % TESS Architect
% \affiliation{Department of Physics and Kavli Institute for Astrophysics and Space Research, Massachusetts Institute of Technology, Cambridge, MA 02139, USA}

\author[0000-0002-6892-6948]{S.~Seager} % TESS Architect
\affiliation{Department of Physics and Kavli Institute for Astrophysics and Space Research, Massachusetts Institute of Technology, Cambridge, MA 02139, USA}
\affiliation{Department of Earth, Atmospheric and Planetary Sciences, Massachusetts Institute of Technology, Cambridge, MA 02139, USA}
\affiliation{Department of Aeronautics and Astronautics, Massachusetts Institute of Technology, Cambridge, MA 02139, USA}

\author[0000-0002-4265-047X]{Joshua~N.~Winn} % TESS Architect
\affiliation{Department of Astrophysical Sciences, Princeton University, 4 Ivy Lane, Princeton, NJ 08544, USA}

\author[0000-0002-4715-9460]{Jon~M.~Jenkins} % TESS Architect
\affiliation{NASA Ames Research Center, Moffett Field, CA 94035, USA}

         %% --------------------------- %%
%% ----- %% ---- TESS CONTRIBUTORS ---- %% ----- %%
         %% --------------------------- %%
%% POC (3)
% \author{John~P.~Doty}% <jpd@noqsi.com>
% \affiliation{Noqsi Aerospace Ltd., 15 Blanchard Avenue, Billerica, MA 01821, USA}
\author[0000-0001-7139-2724]{Thomas~Barclay} %<thomas.barclay@nasa.gov>
\affiliation{NASA Goddard Space Flight Center, 8800 Greenbelt Road, Greenbelt, MD 20771, USA}
\affiliation{University of Maryland, Baltimore County, 1000 Hilltop Circle, Baltimore, MD 21250, USA}
\author[0000-0002-3555-8464]{David Watanabe}% <DavidWatanabeAstronomy@gmail.com>
\affiliation{Planetary Discoveries in Fredericksburg, VA 22405, USA}

%% TSO (2)
\author[0000-0001-6588-9574]{Karen~A.~Collins}%<karenacollins@outlook.com>
\affiliation{Center for Astrophysics ${\rm \mid}$ Harvard {\rm \&} Smithsonian, 60 Garden Street, Cambridge, MA 02138, USA}
\author[0000-0003-3773-5142]{Jason~D.~Eastman}%<jason.eastman@cfa.harvard.edu>
\affiliation{Center for Astrophysics ${\rm \mid}$ Harvard {\rm \&} Smithsonian, 60 Garden Street, Cambridge, MA 02138, USA}

%% SPOC (1)
\author[0000-0002-8219-9505]{Eric~B.~Ting}% <eric.b.ting@nasa.gov>
\affiliation{NASA Ames Research Center, Moffett Field, CA 94035, USA}

%% --------------------------- %%

% \author{\cmc{(Author order is not final, simply a list for the time being)}}

%% --------------------------------------------------- %%
\begin{abstract}
    Large-scale exoplanet surveys like the TESS mission are powerful tools for discovering large numbers of exoplanet candidates.  
    Single-transit events are commonplace within the resulting candidate list due to the unavoidable limitation of observing baseline.  These single-transit planets often remain unverified due to 
    their unknown orbital period and consequent difficulty in scheduling follow up observations.  
    In some cases, radial velocity (RV) follow up can constrain the period enough to enable a future targeted transit detection.  We present the confirmation of one such planet: TOI-2010\,b.  
    Nearly three years of RV coverage determined the period to a level where a broad window search could be undertaken with the Near-Earth Object Surveillance Satellite (NEOSSat), detecting an additional transit.  
    An additional detection in a much later TESS sector solidified our final parameter estimation.  
    We find TOI-2010\,b to be a Jovian planet 
    ($M_P = 1.29 \ M_{\rm Jup}$, 
    $R_P = 1.05 \ R_{\rm Jup}$)
    on a mildly eccentric orbit
    ($e = 0.21$) 
    with a period of 
    $P = 141.83403$ days. 
    Assuming a simple model with no albedo and perfect heat redistribution, the equilibrium temperature ranges from about 360 K to 450 K from apoastron to periastron. Its wide orbit and bright host star ($V=9.85$) make TOI-2010~b a valuable test-bed for future low-insolation atmospheric analysis.
    
\end{abstract}

%\cmc{(check if these need to be from a specific list)}}

%% --------------------------------------------------- %%
\section{Introduction}

Following its launch in 2018, the Transiting Exoplanet Survey Satellite (TESS) mission \citep{TESS_2015} has discovered many thousands of new exoplanet candidates.  
As per its mission mandate, most of these targets orbit stars bright enough for detailed follow up characterization.  
While TESS's nearly full-sky coverage and bright object target list are undeniably valuable qualities, they do come with drawbacks.  One of the most notable is its limited temporal coverage of a given patch of sky.  
% As TESS's Prime and Extended Missions tiled the northern and southern ecliptic hemispheres, it shifted viewing angle every 25-30 days.
TESS's observational strategy has been to shift its viewing angle every 25--30 days to a new sector.
A portion of the sky experiences field overlap between sectors, but a large fraction ($\sim$63\%) receives only month-long baseline coverage.  
This is obviously detrimental for the detection of planets with orbital periods longer than $\sim$30 days.  
At best, TESS might catch one single transit in these regions before moving on to the next sector.  Returning to the field in subsequent sectors can help, but does not guarantee another transit detection.  
Even catching a second transit detection typically leaves many possibilities for the orbital period depending on how many transits may have occurred during the unobserved time interval \citep{Cooke_etal_2021}.  
Without knowledge of the period, certain intrinsic system parameters remain unobtainable or strongly correlated.  
In particular, the semi-major axis and the period (both of which affect transit duration) are largely degenerate.  
As such, determination of stellar irradiation is unavailable.
Attempted measurements of eccentricity and the argument of pericentre are also mostly uninformative.
Lacking a clear picture of the orbital structure makes quantifying the system quite challenging. 
% With a radius determination being one of the few well-characterized properties on a single-transit target, several false positive scenarios remain viable. 
In addition, without strong constraints on the period via multiple transit detections or extensive radial velocity (RV) follow up, scheduling any sort of additional transit-based observations (e.g., transmission/emission spectroscopy, Rossiter-McLaughlin effect, etc.) becomes nearly impossible.

This is unfortunately the fate of most long-period single-transiting planet candidates in the TESS catalogue.  To date, more than 98\% of the 6000+ TESS Objects of Interest (TOIs) with known periods are on orbits shorter than $50$ days
(Exoplanet follow up Observing Program; doi:\dataset[10.26134/ExoFOP5]{https://exofop.ipac.caltech.edu/tess/}), and many of those that have longer reported orbits are poorly characterized and require further verification.
Even if a rough period estimate can be established with RV measurements, the timing uncertainty of future transits grows with each subsequently unobserved transit ($\sigma_{T_n} \propto n\sigma_P $, where $n$ is the number of transits since the period uncertainty, $\sigma _P$, was calculated).
Generally, multiple transit observations are needed to provide tight constraints on the period and keep future timing uncertainties small.

Both transit and RV detection methods suffer observation and detection biases against long-period planets.  In transit surveys, such planets require much longer baseline to capture sufficient events \citep{Beatty_Gaudi_2008}.  
With finite data sets, their folded multi-transit signal-to-noise ratio (SNR) builds more slowly, making shallower transit events especially hard to detect.
Their wide orbital geometries naturally lead to lower transit probabilities, reducing the number of expected events in a given search sample.
RV measurements, which are complementary to transit observations, are also hindered by wider orbits.  The signal amplitude shrinks and it takes longer to cover a full orbit.

Despite and because of these challenges,
there is real value in improving our catalogue of longer-period planets with their cooler equilibrium temperatures
\citep{Fortney_etal_2020}.
Due to these biases and accompanying investment required to study them, long-period planets tend to fall by the wayside and become underrepresented in exoplanet catalogues.
Orbital periods of about 50 days mark a notable boundary in our confirmed planet databases.  
Given that every planet in our own solar system orbits with a period $>50$ days, 
the restriction to our known exoplanet population is quite staggering.
By confirming and cataloguing these wide-orbit planets we build up our understanding of the physical and orbital characteristics within this sparsely measured population.

Though they are few in number, our solar system giants provide detailed data on large cold planets, even allowing \emph{in-situ} measurements \citep[e.g. the Galileo Entry Probe;][]{Neimann_etal_1998}.  Hot, giant, transiting exoplanets also comprise a high-quality data set due to their large sample size and relative ease of detection.
%Daniel note: "between these observable regimes", I think the main advantage is that cool giants are more directly analogous to the solar system giants and don't exhibit the hot-Jupiter radius anomaly (Miller 2014, Thorngren 2016).  That simplifies modeling of them and is a nice control group for doing models of hot-Jupiter inflation.
%
%
Temperate transiting planets in between these extremes require particular effort to observe due to their adverse observational biases.
However, diligent confirmation studies can still accomplish precise measurements of radius, mass, and orbital structure. 
These studies provide information to better understand the long-period planets as individuals, and as a population.

Another advantage of these cool giants is that they are more directly comparable to our well-studied cold solar system giants in that they do not exhibit the hot-Jupiter radius anomaly \citep{Miller_Fortney_2011,Thorngren_etal_2016}.
The relative simplicity of modeling them acts as a valuable control group for understanding hot-Jupiter inflation.
Though RV surveys have measured masses for many cool/cold ($T_{\rm eq} \lesssim 500$\,$K$)
%\cmc{(quantify temp? NEXA?)}
planets, few of them exhibit transits and few of those have reasonably bright hosts, severely limiting their potential for atmospheric characterization.
% The former arises from our own solar system giants where close proximity allows for very detailed characterization of a small sample.  
% The latter consists of the plentiful hot-Jupiter population for which transit and RV surveys are extremely sensitive.
It is a challenge to create generalized chemical or structural atmospheric models that can span the broad temperature range of the giant planet population  without having a solid testing ground in the intermediate range \citep{Gao_etal_2021}.
Chasing down the longest-period targets in the TESS sample helps bridge this gap
\citep[e.g.][]{Dalba_etal_2022}.
%% Chemistry
In terms of atmospheric chemistry, the cooler atmospheres may contain disequilibrium by-products that would serve as valuable probes of atmospheric physics \citep{Fortney_etal_2020}.  
Spectroscopic endeavours can use these lower-insolation targets to tease apart the composition transition between the very cold and very hot giant planet atmospheres.

% \finalchange{(Maybe some of this atmospheric stuff should be in discussion?) }

Stellar insolation can also have many complex effects on a planet's atmosphere.
The question of X-ray and ultraviolet (XUV)-driven mass loss frequently arises in the context of the super-Earth and sub-Neptune populations
\citep{Owen_Wu_2013,Dong_etal_2017,Mordasini_2020}.
Irradiation levels are also important for general structure and evolution models as well as atmospheric circulation and photochemistry in cool planets \citep{Horst_etal_2018}.  
Insolation ought to push the radiative--convective boundary deeper, but also seems to drive the radius anomaly which pushes the boundary back up along with the planet radius
\citep{Thorngren_etal_2019}.
Given the inverse-square law of stellar irradiation,
wider-orbit planets will be significantly less affected by XUV-driven mass loss, preserving more of their primordial composition.
Building a sample of planets with reduced insolation will help with the creation of more broadly applicable planetary models.

%% Dynamics
Long-period planets also provide test cases for system dynamics.
Models describing the formation and migration processes thought to be responsible for the hot-Jupiter population are generally of two categories: 
protoplanetary disk torques 
\citep{Goldreich_Tremaine_1980,Lin_Papaloizou_1986,Ward_1997,Baruteau_etal_2014}
or high-eccentricity migration
\citep{Rasio_Ford_1996,Wu_Murray_2003,Nagasawa_etal_2008,Wu_Lithwick_2011}.
With the end products often being very similar, knowledge of intermediate-separation transiting planets, their companions, and their environments will help distinguish between these types of models.
Transit and RV surveys are always biased against long-period planets, so filling out the eccentricity distribution of long-period giants still requires additional effort
\citep[e.g.][]{Dalba_GOTEM_II}.
Obliquity measurements (using either the RM effect or doppler tomography) have also been done almost exclusively on short-period giants thus far.

% inflated atmospheres (structure theories)
%Daniel - This paragraph seems a little vague.  Insolation pushes the RCB down but also seems to power the radius anomaly (which pushes it right back up, along with the radius).  As far as mass-loss goes, the real issue is the XUV flux rather than the bolometric flux, and these are related in complicated ways, especially with age.  So one thing that can be said about further-out planets is that XUV-driven mass-loss won't be an issue even for sub-Neptunes, preserving the primordial composition most of the time.

TOI-2010 (details in Table~\ref{tab:host_info}) was flagged as containing a transiting planet candidate after a single transit was detected on 2019 August 16. 
% 2019-08-16.
The transiting body, designated TOI-2010.01, did not exhibit a retransit in the remainder of TESS's primary mission.
In this manuscript we confirm and characterize its planetary nature, and so we will hereafter refer to it by the designation TOI-2010\,b in accordance with standard planetary nomenclature.
Its initial single-transit status inspired an intensive RV campaign which constrained the period well enough to make feasible a photometric search for a second transit.  
While both of these efforts were successful, a much later TESS sector fortuitously revealed an additional transit during the late stages of this manuscript preparation, confirming our findings and providing even tighter parameter determination.

We present the various data and observations that contributed to this planet confirmation in Section~\ref{Sec:Data}.  
We then describe the various analyses carried out to characterize the star, planet, and system as a whole in Section~\ref{Sec:Analysis}. 
Sections~\ref{Sec:Results} and \ref{Sec:Discussion} contain descriptions and discussion of our findings, respectively.  
Finally, a brief summary of the entire study is presented in Section~\ref{Sec:Summary}.

%%% HOST STAR INFO TABLE
% \startlongtable
\begin{deluxetable}{lcc}
\tabletypesize{\scriptsize}
\tablecaption{Host Star Information \label{tab:host_info}}
\tablehead{%
  \colhead{Parameter} & 
  \colhead{Value} &
  \colhead{Source} }
  \startdata
TESS ID & TOI-2010 & [1] \\
TIC ID    & TIC 26547036 & [1] \\
Gaia ID   & 2136815881249993600 & [2] \\
$\alpha$  & 19$^{\rm h}$28$^{\rm m}$40.07$^{\rm s}$ & [2] \\
$\delta$  & +53$^{\rm d}$29$^{\rm m}$14.53$^{\rm s}$ & [2] \\
$m_B$   & 10.48 & [3] \\
$m_V$   &  9.85 & [3] \\
$m_G$   &  9.70 & [2] \\
$m_J$   &  8.66 & [4] \\
$m_H$   &  8.34 & [4] \\
$m_K$   &  8.28 & [4] \\
Spectral type   & F0 & [5] \\
Parallax [mas] &  9.2219 $\pm$ 0.0107 & [2] \\
\enddata
% \tablenotetext{}{Note: uncertainties in these values are similar to }
\tablenotetext{}{%
[1] ExoFOP; doi:\dataset[10.26134/ExoFOP5]{https://exofop.ipac.caltech.edu/tess/} \newline
[2] \citet{Gaia_EDR3_2020} \newline
[3] \citet{Hog_etal_2000}  \newline
[4] \citet{Cutri_etal_2003} \newline
[5] Simbad; doi:\dataset[10.17616/R39W29]{http://doi.org/10.17616/R39W29} \newline
}
\end{deluxetable}

% B 10.48 [0.04] D 2000A&A...355L..27H Hog_etal_2000
% V 9.85 [0.03] D 2000A&A...355L..27H Hog_etal_2000
% G 9.700555 [0.002766] C 2020yCat.1350....0G Gaia_EDR3_2020
% J 8.656 [0.026] C 2003yCat.2246....0C Cutri_etal_2003
% H 8.344 [0.017] C 2003yCat.2246....0C Cutri_etal_2003
% K 8.281 [0.016] C 2003yCat.2246....0C Cutri_etal_2003

%% --------------------------------------------------- %%
\section{Data and Observations}\label{Sec:Data}

Numerous observations of this planet candidate were made by the TESS follow up Observing Program (TFOP).  Some of them are overlapping in their coverage or scope, and many were intermediate steps of target validation used to green-light more intensive observations.
We list all the contributions in the subsections below for completeness and recognition, but note that not every data set is included in the analysis that follows.

%% -------------------------------------- %%
% \cmc{
% \subsection{Early vetting and characterization}
% }
% \cmc{\begin{itemize}
%     \item TRES spectra useful for ruling out SB2 or other FP. Not useful for RVs or spectroscopic characterization of star (May not need to even mention)
%     \item Keck/HIRES is superior for characterization
%     \item 
% \end{itemize} }

% \begin{figure*}[t]
% 	\centering
% % 	\includegraphics[width=0.47\textwidth] 
% 	\includegraphics[width=0.98\textwidth]
%     {Pauls_figures/rv_v2.pdf}
	
%     \caption{RV coverage of 110 measurements spanning $\sim$2.7 years uniformly samples the full phase of the planet's orbit.
%     %
%     %\cmc{(Check to make sure it's not a known wide-binary.  Gaia? Visier? Simbad? (can look for WDS - Washington Double Star tag) , could look at gaia data and try matching parallax and PM of nearby stars    Note: VisieR seems to show a WDS entry, but doesn't give much info on the secondary except for a 1.6" separation.  Check the ShARKS data once it comes in for this companion.)}
%         \label{fig:RVs}}
% \end{figure*}

%% -------------------------------------- %%
\subsection{Discovery and Sky-Monitoring Photometry} \label{ssec:sky_monitoring_phot}

%% ---------------- %%
\subsubsection{TESS}

%%% VSG found it around Oct 10, and shared with TSTPC around Oct 15, 2019

In the early stages of the preparation of this manuscript, TOI-2010 had been observed at 2 minute cadence for Sectors 14, 15, 16, and 40 and the image data were reduced and analyzed by the Science Processing Operations Center \citep[SPOC;][]{SPOC_2016} at NASA Ames Research Center.  
A single transit was detected at the beginning of Sector~15 using an adaptive, wavelet-based matched filter 
\citep{Jenkins_2002,
Jenkins_etal_2010,
Jenkins_etal_2020}
on 
% 21 September 2019.
2019 September 21.
The signal was also independently discovered by the Visual Survey Group \citep{Kristiansen_etal_2022} around the same time and forwarded to the attention of the TESS Single Transit Planet Candidate (TSTPC) working group for follow up.
It was alerted by the TESS Science Office as a Community TOI (CTOI) on 17 June 2020.
This single transit event, though unambiguous due to its high SNR, posed a validation challenge due to its lack of period constraint.

%% Late-game transit detection
During the late stages of manuscript preparation, TESS reobserved TOI-2010 in Sectors 54, 55, and 56 as part of its extended missions.
An additional transit was detected in the Sector 56 light curve and included in our final analysis.
A search by the SPOC of Sectors 14--56 reported a difference image centroiding result \citep{Twicken_etal_2018} constraining the host star location to within $2\farcs54 \pm 2\farcs9$ of the difference image centroid.  
This is in agreement with the low contamination by reported nearby Gaia stars.

%% Getting the TESS data files (MAST)
We acquired the Pre-search Data Conditioning Simple Aperture Photometry 
\citep[PDCSAP;][]{Stumpe_etal_2012,Stumpe_etal_2014,Smith_etal_2012}
flux data from the Milkuski Archive for Space Telescopes (MAST) for the 2 minute Sector 14, 15, 16, and 40 data (doi:\dataset[10.17909/t9-nmc8-f686]{https://dx.doi.org/10.17909/t9-nmc8-f686}) and the 20\,s Sector 54, 55, and 56 data (doi:\dataset[10.17909/t9-st5g-3177]{https://dx.doi.org/10.17909/t9-st5g-3177}).

\begin{figure*}[t]
	\centering
	\begin{tabular}{cc}
    \includegraphics[width=0.98\textwidth]{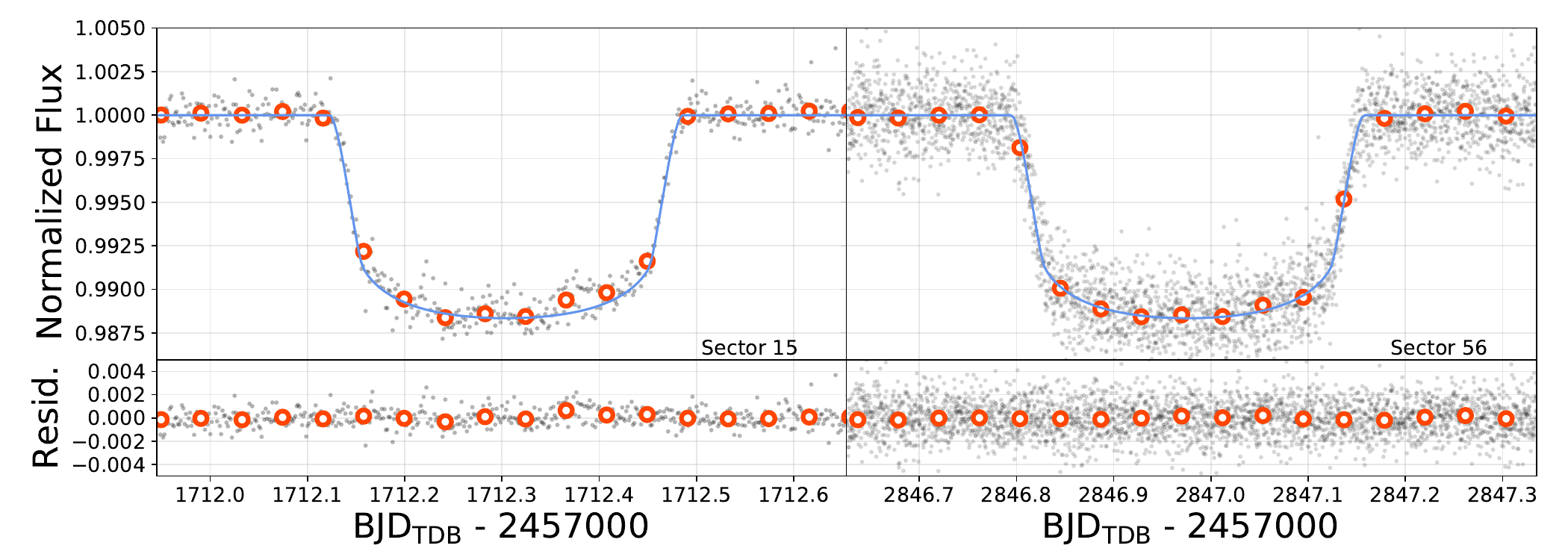}
	\end{tabular}
    \caption{
    PDCSAP data of the TESS transit detection in Sector 15 (left) and Sector 56 (right). Sector 15 was imaged with a 120\,s cadence, while Sector 56 included a 20\,s cadence. All bins are 60 minutes.
        \label{fig:TESS_transit}}
\end{figure*}

Before applying any light curve fitting, we use the {\tt lightkurve} \citep{Lightkurve_2018} software package's built-in {\tt flatten} routine to remove any remaining PDCSAP variability.  
This applies a Savitzky--Golay filter \citep{Savitzky_Golay_1964} to the light curves, fitting a low-order polynomial to a rolling subset of the data to remove low-frequency trends.  The transit regions were masked during this process and the subset windows were chosen to be longer than the transit duration.
We apply the same {\tt flatten} routine to
the 20\,s cadence Sector 56 TESS data,
applying an additional rolling sigma-clip routine (3$\sigma$ from the median in a window of $\pm$100~minutes) to remove outliers.
The portions of the flattened light curve containing the transits and used for analysis are shown in Figure~\ref{fig:TESS_transit}.

With the current Year 5 plan for the TESS mission, there are no scheduled visits to this region of the sky after Sector 56.

%% ------------------------- %%
\subsubsection{WASP}

The field containing TOI-2010 was observed by the Wide Angle Search for Planets (WASP) transit-search survey 
% \citep{2006PASP..118.1407P} 
\citep{WASP_paper}
from 2008 to 2010. In each year the observing season spanned $\sim$\,130 nights, with the SuperWASP-North camera array observing the field on clear nights with a typical 15 minute cadence.
A total of 32\,000 photometric data points were obtained using 200\,mm, $f$/1.8 Canon lenses backed by 2k$\times$2k CCDs. TOI-2010 is the only bright star in the 48\arcsec\ extraction aperture.

The WASP data are dominated by systematics and red noise. 
While the transit depth is likely sufficient to show up in the WASP light curves, the survey mission relies on multiple repeated events to distinguish transits from noise.
As a single-transit target at the time, WASP did not detect any events on the target. 
Even with the benefit of hindsight and a firm orbital ephemeris, the WASP coverage only overlaps with one predicted transit.
The 4\,hr span of data lies in the middle of a 8.7\,hr transit, and the light curve shows no convincing transit-like features.

%%% Potential figure:
% \begin{figure}
% \centering
%     \includegraphics[width=0.47\textwidth]
%     {WASP_modulations.pdf}
    	
%     \caption{Periodograms of the WASP lightcurves of TOI-2010. The top panel combines the data from all 3 years. The horizontal lines show the estimated 10\%\- and 1\%-likelihood false-alarm levels. At right are the data folded on the 20-d period.}
% \label{fig:wasp}
% \end{figure}

%% -------------------------------------- %%
\subsection{Candidate Vetting} \label{ssec:vetting_obs}

Once established as a TOI, a number of vetting observations were undertaken. They were used to search for false positive indications, and to assess the target's suitability for further follow up observations.

%%% SPECTROSCOPY TABLE
% \startlongtable
\begin{deluxetable}{lcccc}\label{tab:spec_params}
\tablecaption{Stellar Parameters from Independent Spectral Instruments/Measurements}
\tablehead{\colhead{Parameter} & \colhead{HIRES} & \colhead{NRES} & \colhead{TRES} & \colhead{Units} }
\startdata
$T_{\rm eff}$ & $5917\pm75$ & $5860 \pm 100$ & $5795 \pm 50$ & K \\
$\log g$ & $4.412^{+0.023}_{-0.026}$ & $4.5\pm0.1$ & $4.42\pm0.10$ &  \\
$[{\rm Fe/H}]$ & $0.169^{+0.055}_{-0.056}$ & $0.23\pm0.06$ & $0.22\pm0.08$ &  \\
$M_\star$ & $1.107^{+0.050}_{-0.057}$ & $1.139\pm0.049$ & -- & $M_\odot$ \\
$R_\star$ & $1.084^{+0.028}_{-0.027}$ & $1.106\pm0.074$ & -- & $R_\odot$ \\
% $v\sin i_\star$ & $1.3 \pm 1.0$ & $3.5 \pm 0.9$ & $4.3 \pm 0.5$ & km s$^{-1}$ \\
$v\sin i_\star$ & $<2.3$ & $<4.4$ & $<4.8$ & km s$^{-1}$ \\
\enddata
\tablenotetext{}{
Note: $v\sin i_\star$ values become challenging to constrain when at the few kilometres per second level as many line-broadening mechanisms are simultaneously at play on this scale (e.g. macroturbulence). We therefore treat the spectral estimates as upper limits.
% \cmc{(Andrew Howard notes that HIRES $v\sin i$ is an upper limit and suggests just posting $<2$ km/s.  Same story about other broadening mechanisms at this low level.)}
}
\end{deluxetable}

%% LCO / NRES
%% EXOFAST fit has similar or tighter errors on each
% Teff=5860+/-100     % agrees with EXOFAST
% logg=4.5+/-0.1      % agrees with EXOFAST
% [Fe/H]=0.23+/-0.06  % mostly agrees with EXOFAST
% Mstar=1.139+/-0.049 % agrees with EXOFAST
% Rstar=1.106+/-0.074 % agrees with EXOFAST
%
% vsini=3.48+/-0.87 % is not reported by EXOFAST

%% FLWO / TRES
%%% Teff has tighter errors on TRES
% Teff=5795+/-50K   % just agree at 1-sigma 
% logg=4.42+/-0.10  % agree
% [m/H]=0.22+/-0.08 % agree
% SNRe=30.5 % ??

% vsini=4.3+/-0.5 km/s

\subsubsection{Keck/HIRES Spectra}\label{sssec:HIRES}

We obtained a spectrum of TOI-2010 with the High Resolution Echelle Spectrometer \citep[HIRES;][]{HIRES_1994} on the Keck I telescope at W. M. Keck Observatory to explore false positive explanations for the single-transit event, to assess the quality of the host as a target for Doppler spectroscopy, and to conduct a basic spectral characterization of the host. 
Initial processing of the spectrum with \texttt{SpecMatch-Emp} \citep[``Emp" indicating the ``Empirical" flavour of the code;][]{SpecMatch_Empirical} determined the stellar parameters.
The results, along with the star's bright
% $V=9.9$ 
magnitude indicated that it would likely be a suitable target for Doppler spectroscopy.
%via the APF \citep{APF_2015}. 
%
% By processing the spectrum with \texttt{SpecMatch-Emp} \citep[``Emp" indicating the ``Empirical" flavour of the code:][]{SpecMatch_Empirical}, we inferred a stellar effective temperature of 
% $T_{\rm eff} = 5878\pm100$~K, 
% a surface gravity of 
% $\log{g}= 4.40\pm0.01$, 
% an iron abundance of 
% [Fe/H] = $0.19\pm0.06$, 
% and a stellar rotational velocity of 
% $v\sin{i} = 1.3\pm1.0$~km~s$^{-1}$. 
%

This Keck/HIRES measurement was taken under excellent seeing conditions and produced a spectrum with SNR $\sim$ 200.  Given this data quality it was used as the template spectrum with which the Levy RV measurements were extracted (see Section~\ref{ssec:RV_obs}).  Similarly, we favour the Keck/HIRES extracted stellar parameters over those from LCOGT/NRES and FLWO/TRES (Sections~\ref{sssec:LCOGT_NRES} and \ref{sssec:FLWO_TRES}) due to the quality of the spectrum, though we note the close agreement of most parameters.  
Table~\ref{tab:spec_params} provides a comparison of these stellar parameters.

% \pdc{[Allyson says TRES gives vsini=4.3 $\pm$ 0.5 km/s. 
% NRES seems to give 3.5 $\pm$ 0.9 km/s.
% These seem wildly different (and more precise?) than what you have here from HIRES (1.3 $\pm$ 1.0 km/s...).  This has pretty strong bearing on RM measure.]}
% \cmc{We could report all of these values. I'm not an expert on spectral classification, but my understanding is that for slowly rotating stars, say below 5 km/s, vsini is actually hard to measure accurately, because it is on the same order as (or less important than), turbulence, thermal broadening, pressure broadening, and broadening caused by the profile of the instrument itself. To make things more confusing, SpecMatch was also run on the lower SNR vetting spectrum. It returned 3.1 $\pm$ 1.0 km/s!! We could mention this as well and give the mean and spread of each of these as our best guess for the vsini.}

\subsubsection{LCOGT/NRES Spectra}\label{sssec:LCOGT_NRES}

% from Markus Rabus
We scheduled spectroscopic observations for TOI-2010 on the
Las Cumbres Observatory Global Telescope (LCOGT; \citet{LCO_paper})
Network of Robotic Echelle Spectrographs \citep[NRES;][]{Siverd2018}. NRES comprises four identical echelle spectrographs in different observatories, covering a range of longitudes in the Northern and Southern Hemispheres. 
The resolving power of the echelle spectrographs is $R\sim 53,000$ covering the wavelength range 3900--8600 \AA. 
We obtained four good-quality (SNR~18--56) spectra with the NRES unit at the Wise Observatory between 
% June 19, 2020 
2020 June 19 and 28.
% and 
% June 28, 2020. 
We used the 
% \texttt{CERES} pipeline \citep{Brahm2017} 
{\tt BANZAI-NRES} pipeline \citep{BANZAI-NRES}  
to reduce the spectra and extract RVs, and the {\tt SpecMatch-Synthetic} code for the stellar parameterization \citep{SpecMatch_Synthetic_thesis,SpecMatch_2017}.

While the NRES observations provided helpful early vetting of the system, we have chosen to exclude the four RV measurements from the analysis due to their much lower precision ($>$20\,m\,s$^{-1}$). 
The derived stellar parameters are generally in close agreement with the Keck/HIRES values (Table~\ref{tab:spec_params}).
% Stellar parameters (with one exception) came out in close agreement with the Keck/HIRES values. 
% The one discrepant parameter was the rotational value of $v\sin i_\star = 3.48\pm0.87$ km/s.
% We use the Keck/HIRES values due to the superior SNR and spectral coverage.

% \cmc{Seems to have one LCO spectrum loaded to TFOP under the name of Markus Rabus. Resolution of 53000 over 380 to 860 nm.
% %
% Provide a PNG image file that shows some of the spectrum and a fitted model. Provide estimates of $T_{eff}, \log g, {\rm Fe/H}, v\sin i, M_\star, R_\star$ with uncertainties}

%% EXOFAST fit has similar or tighter errors on each
% Teff=5860+/-100     % agrees with EXOFAST
% logg=4.5+/-0.1      % agrees with EXOFAST
% [Fe/H]=0.23+/-0.06  % mostly agrees with EXOFAST
% Mstar=1.139+/-0.049 % agrees with EXOFAST
% Rstar=1.106+/-0.074 % agrees with EXOFAST
%
% vsini=3.48+/-0.87 % is not reported by EXOFAST

\subsubsection{FLWO/TRES Spectra}\label{sssec:FLWO_TRES}

% From Alysson Bieryla:
Three reconnaissance spectra of TOI-2010 were obtained on 
% UT2020-07-09, 
% UT2020-07-18, and 
% UT2020-07-27 
% July 9, 18, and 27 of 2020
2020 July 9, 18, and 27
with the Tillinghast Reflector Echelle Spectrograph 
\citep[TRES;][]{gaborthesis}.
TRES is an optical (390--910\,nm) spectrograph with a resolving power of $R\sim 44\,000$ mounted on the 1.5\,m Tillinghast Reflector telescope at the Fred Lawrence Whipple Observatory (FLWO).
The spectra, with SNR in the range of 25--35, were extracted using the TRES standard pipeline 
\citep{buchhave2010}
and the stellar parameters were derived using the Stellar Parameter Classification 
\citep[SPC;][]{buchhave2012,buchhave2014}
tool. SPC cross correlates the observed spectra against a grid of synthetic spectra based on Kurucz atmosphere models 
\citep{kurucz1992}
deriving stellar effective temperature, surface gravity, metallicity, and rotational velocity. 

In the same sense as the NRES spectra described above, these data were useful in the early classification of the star and ruling out of false positives, enabling more detailed measurements to be carried out.  The FLWO/TRES stellar parameters are also generally in close agreement with the Keck/HIRES analysis (Table~\ref{tab:spec_params}).
% with the exception of 
% $v\sin i_\star = 4.3 \pm 0.5$ km/s.
% \cmc{ This determined $v\sin i_\star$ value does not account for macroturbulence and so might also be treated as an upper limit.}
% Again, we adopt the Keck/HIRES values.

% \cmc{Paul says the TRES spectra were useful for checking for SB2 or other FP before this was even a TOI.  He says we don't need to include or even mention TRES, but it would be nice to include Allyson (Dave will be included by default).}

% \cmc{Three TRES observations loaded to TFOP from July 2020. Resolution of 44000 over 3850 to 9096 angstroms. Under the name of Allyson Bieryla.
% %
% Uploaded three classification PDFs.  They provide estimates of $T_{eff}, \log g, {\rm [m/H]}, V_{rot}$ without any uncertainties.}

% \cmc{Allyson contacted me (2022-09-26) with TRES parameters and error bars.  Can mention, but values are similar to what we already have, so we can just display Keck values.}

%%% Teff has tighter errors on TRES
% Teff=5795+/-50K   % just agree at 1-sigma 
% logg=4.42+/-0.10  % agree
% [m/H]=0.22+/-0.08 % agree
% SNRe=30.5 % ??

% vsini=4.3+/-0.5 km/s

%% ------------------------- %%
\subsubsection{Gemini-N/`Alopeke Imaging}

\begin{figure}[t]
	\centering
	\includegraphics[width=0.47\textwidth]
    {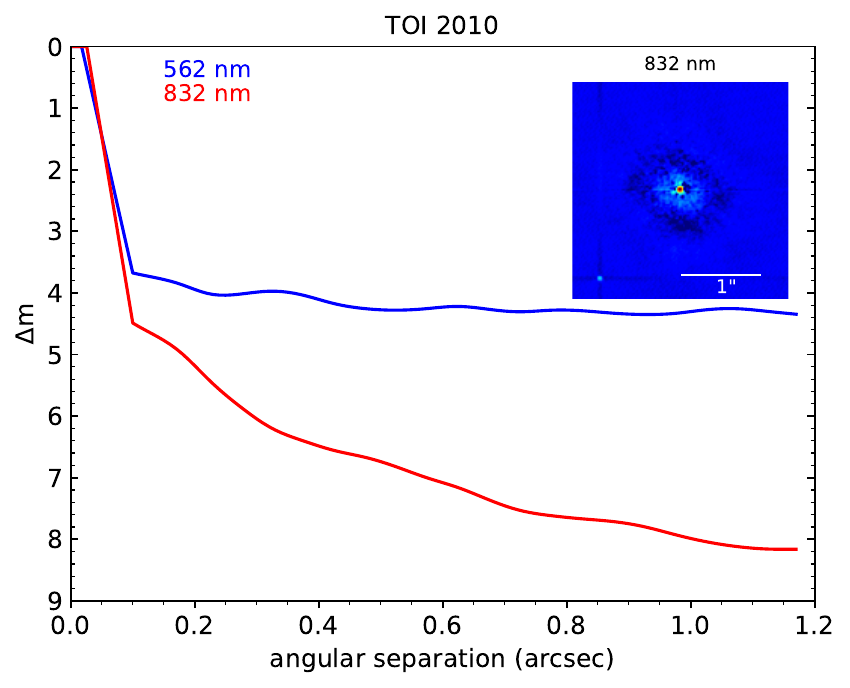}
	
    \caption{
    Contrast curves from Gemini-N/`Alopeke speckle image.  
    Curves show the $5\sigma$ contrast limit.
    The faint source in the lower left of the inset image is a previously unresolved neighbour star not present in the Gaia DR3 catalogue.
        \label{fig:Gemini_speckle}}
\end{figure}

Using the `Alopeke instrument mounted on the Gemini-North telescope we acquired high-contrast imaging of TOI-2010 on 
% 7 June, 2020 
2020 June 7
(Program ID: GN-2020A-Q-132).  
This observation was a part of the exoplanet follow up campaign by \citet{Howell_etal_2021}.  With `Alopeke's design, it can simultaneously capture imagery at both 562\,nm and 832\,nm. 
The resulting images were processed using the pipeline of \citet{Howell_etal_2011}, and the resulting contrast curves are shown in Figure~\ref{fig:Gemini_speckle}.  
Due to the clearly superior performance of the 832\,nm filter, we use the red contrast curve for all analyses (e.g. Section \ref{ssec:grid-search}).

One previously unknown nearby source was detected, seen in the lower left corner of the Figure~\ref{fig:Gemini_speckle} inset.  This object is separated by 1\farcs5 at a position angle of 138$^\circ$ east of north.  Brightness uncertainty on this neighbour is somewhat elevated as it lies outside the speckle correlation radius of $\sim$1\farcs2.  Speckle decorrelation begins to set in beyond this separation when the rays do not pass through the same atmospheric path. Our photometric estimate of the source places it at a $\Delta {\rm mag} = 5 \pm 0.5$.

This source is not to be confused with the neighbouring Gaia star discussed further in Section~\ref{Sec:Results}. 
The Gaia source lies at a separation of 1\farcs9 and a position angle of 33$^\circ$ east of north, beyond the field of view of our `Alopeke image. 
The 1\farcs5 source found in our `Alopeke image does not appear in Gaia's DR2 or DR3 catalogues.

\begin{figure*}[t]
	\centering
	% Trim:  Left  Bottom  Right   Top
	\includegraphics[width=0.98\textwidth,
	trim=0 10 0 0,clip]
    {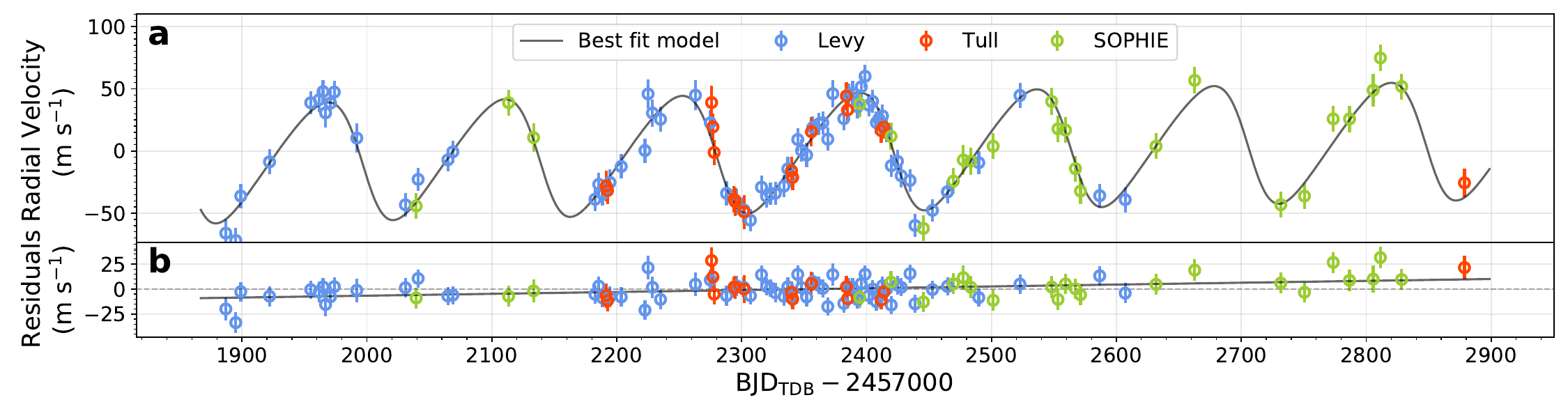}
	\includegraphics[width=0.98\textwidth,
	trim=0 0 10 5,clip]
    {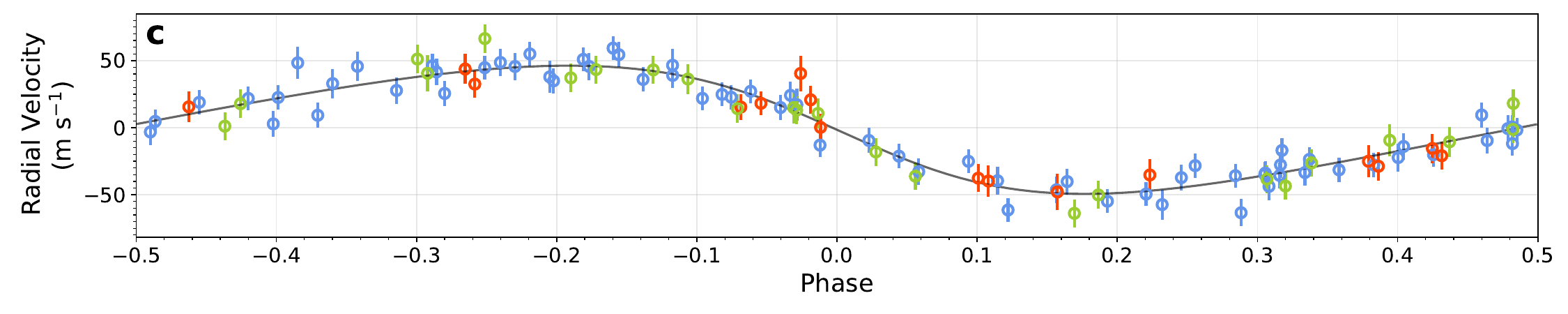}

    \caption{
    RV coverage of 110 measurements spanning $\sim$2.7 yr uniformly samples the full phase of the planet's orbit.
    A small residual acceleration remains after the removal of the planet's Keplerian signal. 
        \label{fig:RVs}}
\end{figure*}

%% -------------------------------------- %%
\subsection{Radial Velocities} \label{ssec:RV_obs}

% The various reconnaissance spectra were able to determine TOI-2010's suitability for RV follow up.
%
The reconnaissance spectra from Keck/HIRES, LCOGT/NRES, and FLWO/TRES were able to place sufficient constraints on the stellar parameters to identify TOI-2010 as a suitable candidate for precise RV measurements.
We collected a total of 110 RV measurements (Figure~\ref{fig:RVs}, values in the Appendix) to make up our combined RV data set.  
These measurements come from three separate instruments, span 992 days, and uniformly sample the phase space of the 142 day periodic signal that stands out in the data.

%% ---------------- %%
\subsubsection{Levy}\label{sssec:Levy}

In February of 2020 (BJD 2458887),
we began to gather spectra on the target for RV measurements.
We started with the Levy spectrograph installed on the 2.4\,m Automated Planet Finder (APF) telescope at Lick Observatory in California, acquiring 70 spectra over a 2 year period, carried out by the dynamic queue scheduler
\citep{APF_2015}. 
The Levy spectrograph is a high-resolution ($R\sim114\,000$) slit-fed optical echelle spectrometer \citep{Radovan_etal_2010} that has previously been used to refine the orbital period and mass of single-transit planet candidates identified by TESS \citep[e.g.,][]{Dalba_etal_2022}. 
We gathered spectra with exposure times of 20--25 minutes (mostly 25), achieving signal-to-noise ratio (SNR) values of 50--100 at around 550\,nm.
An iodine cell in the light path allows for wavelength calibration and the forward modeling of the stellar RV for each spectrum \citep{Butler_etal_1996,Fulton_etal_2015}. 
This forward modeling process relies on having a high-SNR spectrum that is used as a template. 
The HIRES spectrum described in Section~\ref{sssec:HIRES}, which had a SNR of roughly 200, was used to create this template spectrum for the extraction of the Levy RVs which were obtained with uncertainties of 4--7\,m\,s$^{-1}$.

We look for correlations in the $\log R'_{HK}$ activity index (computed from the S-index using {\tt PyAstronomy} routines) with RVs to determine if stellar activity may be biasing the measurements. 
We determine a correlation coefficient of 
$0.10 \pm 0.05$ 
and a $p$-value of
$0.43 \pm 0.23$, 
indicating no evidence of correlation.

%% ---------------- %%
\subsubsection{Tull}

%% from William (Bill) Cochran
We also gathered high-precision RV observations at the McDonald Observatory using the Tull coud\'{e} spectrometer 2 (TS2) on the 2.7\,m Harlan J.  Smith Telescope \citep{Tull_etal_1995}.   
This cross-dispersed echelle white-pupil spectrometer was used in its ``TS23" mode (indicating the third focus)
with an entrance slit of $1\farcs2 \times 8\farcs2$, giving a spectral resolving power of $R = 60\,000$ over most of the visible spectrum.  
A temperature-stabilized $I_2$ gas absorption cell in front of the spectrograph entrance aperture provided the velocity calibration.   
An exposure meter recorded the time series of flux entering the spectrograph, enabling us to compute the flux-weighted barycentric correction.   
A wave front sensor was used for telescope focus to optimize pupil illumination stability and throughput.
We obtained the measurements with 20--30 minute exposures, achieving an SNR per pixel of 62--96 (mean $\sim$ 75).
The spectra are recorded on a $2048 \times 2048$ pixel Tektronix CCD.    
All spectra were reduced and 1D spectra were extracted using standard {\tt IRAF} routines \citep{Tody_1993,Tody_1986}. 
In all, a total of 16 spectra of TOI-2010 were obtained between 
2020 December 8   
%% 2020-12-08 01:54:40.118
% 2459191.579631
and 
% 2022 May 21.  %% 2021-05-21 08:59:11.702
% % 2459355.874441.
2022 October 26. 
%% 2022-10-26 03:04:03
% 2459878.627818
RVs were computed using the {\tt AUSTRAL} code \citep{Endl_etal_2000}, resulting in uncertainties of 9--12\,m\,s$^{-1}$.

We conduct a similar activity--RV correlation search as was done with the Levy, resulting in a coefficient of
$0.02 \pm 0.20$
and a $p$-value of
$0.59 \pm 0.26$ for Tull.
Again, there is no indication of RV correlation.

%% ---------------- %%
\subsubsection{SOPHIE}

We started observing TOI-2010
%TIC 26547036 
with the Spectrographe pour l’Observation des Phénomènes des Intérieurs stellaires et des Exoplanètes (SOPHIE) in July of 2020,
% to characterize its transiting object, soon after it was identified as a TESS Single Transit Planet Candidate. 
securing 25 spectroscopic measurements up to September of 2022. SOPHIE is a stabilized \'echelle spectrograph dedicated to high-precision RV measurements in optical wavelengths on the 193\,cm Telescope at the Observatoire de Haute-Provence, France
\citep{Perruchot_etal_2008,Bouchy_etal_2009}. 
We used the SOPHIE high-resolution mode (resolving power $R=75\,000$). 
Depending on the weather conditions, the exposure times ranged from 11 to 30 minutes (typically 18 minutes) and their SNR per pixel at 550\,nm ranged from 24 to 55 (typically 46). 
The corresponding RVs were extracted with the standard SOPHIE pipeline using cross-correlation functions 
\citep{Bouchy_etal_2009}
and including CCD charge transfer inefficiency correction 
\citep{Bouchy_etal_2013}. 
Following the method described, e.g., in 
\citet{Pollacco_etal_2008} and \citet{Hebrard_etal_2008}, 
we estimated and corrected for the moonlight contamination using the second SOPHIE fiber aperture, which is targeted on the sky while the first aperture points toward the star. 
We estimated that four of the 25 spectra were significantly polluted by moonlight; one of which was too contaminated and was excluded.
The other three contaminated measurements were corrected, with corrections below 20\,m\,s$^{-1}$.
Thus our final SOPHIE data set included 24 measurements showing RV uncertainties ranging 3--9\,m\,s$^{-1}$.

SOPHIE $\log R'_{HK}$ activity measures similarly show no correlation with RV values. 
We determine a correlation coefficient of
$0.22 \pm 0.14$
and a $p$-value of 
$0.38 \pm 0.27$.

\begin{figure}[t]
	\centering
    \includegraphics[width=0.47\textwidth]{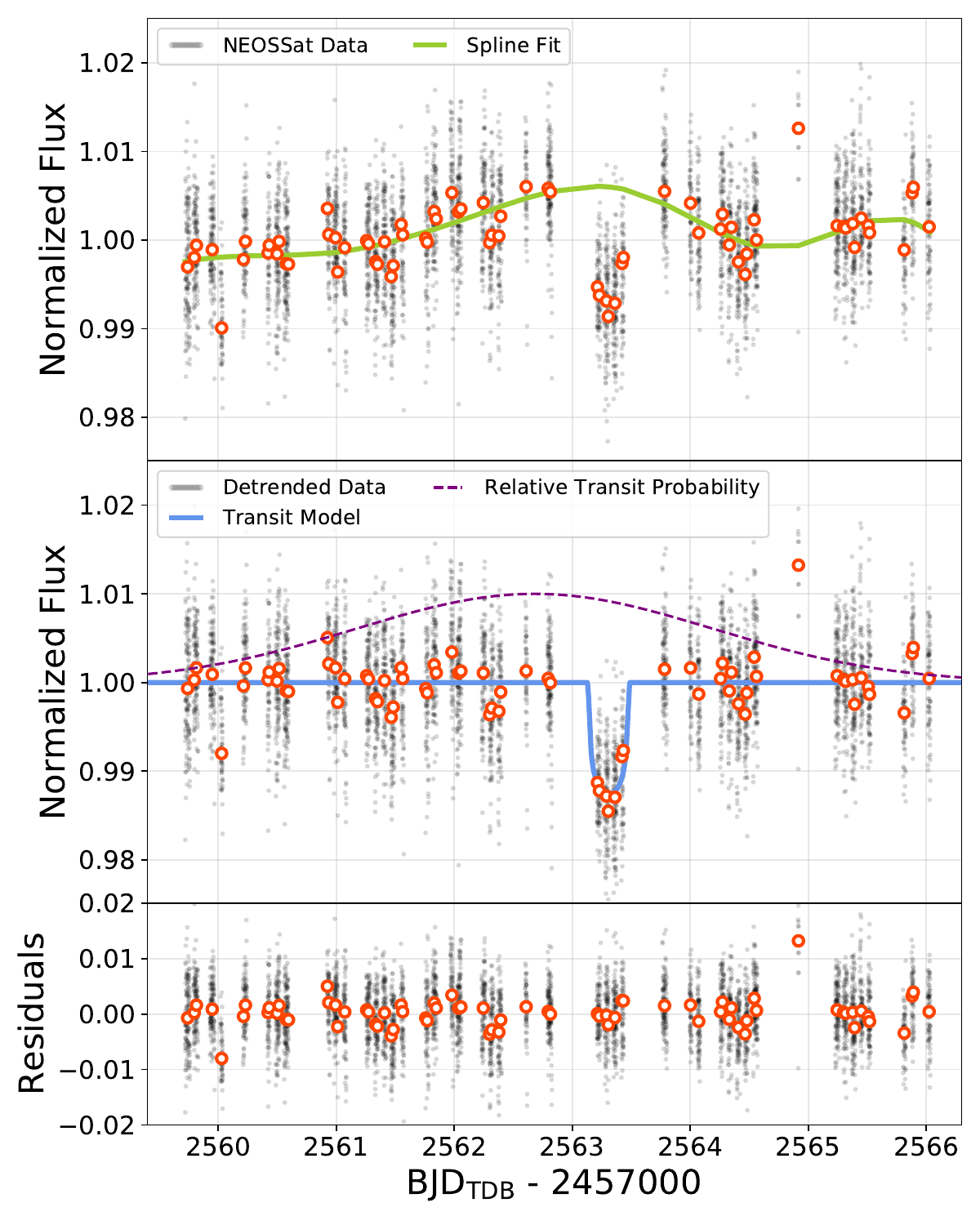}
	
    \caption{
    A week-long observation by NEOSSat. 
    Gaps in the light curve are due to Earth eclipse and other necessary telescope operations. The star was imaged with a cadence of 23\,s while on target.  Displayed bins are 60 minutes.
    \emph{Top:} A \texttt{KEPLERSPLINE} fit to the out-of-transit region to model systematics (discussed in Section~\ref{ssec:transit_modelling}).
    \emph{Middle:} The transit fitted to the corrected light curve.  The dashed purple curve shows relative probability of the expected transit based on the RV-derived period constraints available at the time of observation.
    \emph{Bottom:} Residuals of the transit fit.
    \newline
        \label{fig:NEOSSat_transit}}
\end{figure}

%% -------------------------------------- %%
\subsection{follow up Photometry} \label{ssec:followup_phot_obs}

The collective RV campaign was able to map out a clear planetary signal, but the period uncertainty was only constrained to the order of a few days.  This was insufficient for reliable scheduling of transit observations, so we undertook a few efforts to catch a subsequent transit and fine-tune the period.

%% ------------------------- %%
\subsubsection{GMU}

We observed TOI-2010 with the George Mason University Observatory's 0.8\,m Ritchey--Chretien telescope on the nights of the 
% 21 and 22 of July, 2021 
2021 July 21 and 22
to capture a second transit.  
We imaged in R with an SBIG-16803 CCD with exposure times of 30\,s repeated for a duration of $\sim$5 and 2.5\,hr each night, respectively. 
Both nights were impacted by intermittent clouds, and single measurement precisions of 6.5 and 7.5 ppt were obtained per 30 second exposure. 
Data was reduced and plate-solved using a custom python code {\tt alnitak}\footnote{ \url{https://github.com/oalfaro2/alnitak}} and aperture photometry, reference star selection, and systematic detrending were performed with {\tt AstroImageJ} \citep{AIJ_2017}.

This attempt was prompted due to a predicted transit (from preliminary RV fits) occurring very near the end of TESS Sector 40, and motivated by an absence of TESS coverage in Sector 41.
Unfortunately, no transit was detected on either night.
Given the broad transit timing uncertainty at the time, the narrow available observing windows, and the very long transit duration, the odds of detecting the transit here were quite low.  
Unbeknownst at the time, the transit occurred 1.8 days after the second observation.
These data provided initial constraining power for refining the RV period, but do not benefit the global orbital model. 
As such, they are not included in the modeling of Section~\ref{Sec:Analysis}.

%% ------------------------- %%
\subsubsection{NEOSSat}

The Near-Earth Object Surveillance Satellite (NEOSSat) is a small spacecraft operated jointly by the Canadian Space Agency (CSA) and Defence Research and Development Canada (DRDC).  It has a 15\,cm telescope aperture and is capable of precision relative photometry \citep{NEOSSat_2019}.
NEOSSat's clear-filter effective bandpass is approximately 400--900\,nm.

As the RV data accumulated, preliminary joint fits (see Section~\ref{Sec:Analysis} for details) of the RVs and TESS sector 15 transit revealed a roughly 142 day period, albeit with  broad uncertainties.  
The extended transit duration ($\sim$8.7\,hr), wide timing uncertainty (on the order of a week), and long period made observing a subsequent transit from the ground extremely challenging.  
While NEOSSat had previously proven its capability with short-period exoplanet follow up \citep[e.g.,][]{Fox_Weigert_2022}, TOI-2010\,b marked the first attempt at using the instrument to recover a long and uncertain period.
With its space-based vantage point, NEOSSat has the capability to stare continuously at a target for an extended duration, interrupted only by Earth-eclipse events and technical satellite operations.  
In mid-December of 2021 we employed NEOSSat to observe a $\sim$6 day ($2\sigma$) window around the predicted transit.  The telescope imaged TOI-2010 as continuously as was feasible during this time.

We reduced the raw images and extracted aperture photometry using a custom python pipeline developed for NEOSSat, available on GitHub.\footnote{\url{https://github.com/jasonfrowe/neossat}}
With photometry in hand, we applied a principal component analysis (PCA) procedure to the raw photometry using other in-frame stars as reference to calculate a normalized relative flux light curve of the target.  
The PCA process removes time-varying trends in the photometry that are common across many stars in the frame.
At this point there remained some residual variability for which the PCA could not account.  
This low-frequency variability was removed using a spline fit (discussed in Section~\ref{ssec:transit_modelling}).
Even against this variability the deep transit event was clearly visible roughly 15\,hr after the RV-predicted midpoint (well within the 3\,day, $1\sigma$ timing uncertainty).

This detection by NEOSSat provided the first precise period measurement for TOI-2010\,b, and prompted the preparation of this manuscript.  The much later transit detected in TESS's Sector 56 data agrees completely with the refined period.
The NEOSSat light curve is displayed in Figure~\ref{fig:NEOSSat_transit}.

%% -------------------------------------- %%
%% -------------------------------------- %%
%% -------------------------------------- %%
%% -------------------------------------- %%
%% -------------------------------------- %%

%% This bit is covered in the modeling section
% To flatten the remaining variability, we make a spline fit to the out-of-transit data based on the \texttt{KEPLERSPLINE} of \cmc{Vanderburg \& Johnson (2014)} which was initially designed to handle Kepler-like systematics in long light curves.  This process is displayed in Figure~\ref{fig:NEOSSat_transit}.
%

%% -------------------------------------- %%
% \subsection{High contrast/resolution imaging}

% \cmc{(Without the ShARCS data included in the paper, reconsider the subsection structure here...)}

% %% ------------------------- %%
% \subsubsection{Shane/ShARCS imaging}

% \cmc{---WILL CUT UNLESS THEY RSVP SOON---}
% \newline

% \cmc{NIR Adaptive Optics observation posted on TFOP by Courtney Dressing from 2021-07-19.  
% I don't know if it's better/worse than the `Alopeke imagery.  There is no data uploaded alongside the ``observation" details.  There seem to be two filters. }

% \cmc{(Courtney Dressing will join the paper with her colleague Steven Giacalone.  Steven will provide the contrast curve and a technical blurb about the telescope/instrument.  
% Pinged Steven on Oct 4, Nov 1, Nov 18 2022, and Jan 9 2023.)}

%% --------------------------------------------------- %%
\section{Analysis}% and Global fit of Stellar and Planetary Parameters}
\label{Sec:Analysis}

%% Set up the scenario
As the RV campaign progressed, we made preliminary fits using the TESS light curve and the available RV data to place initial constraints on the orbital period.
This allowed us refine the ephemeris enough to plan our follow up search for a subsequent transit event. 
Once they became available, the additional transit detections (NEOSSat and TESS Sector 56) allowed for much more precise period determination.

%% EXOFASTv2
For our final global fit, we use the IDL software package \texttt{EXOFASTv2} \citep{Eastman:2019}.  
\texttt{EXOFASTv2} provides an integrated framework to jointly analyze multiple exoplanet data sets.
Drawing from the IDL astronomy library \citep{Landsman_1993}, it simultaneously fits for wide ranges of stellar, planetary, orbital, and instrumental parameters in a self-consistent manner that leverages the rich complementarity of modern data sets.

The details listed below in Sections~\ref{ssec:SED/MIST}--\ref{ssec:transit_modelling} pertain to the final fit, including archival SED measurements; RVs from Levy, Tull, and SOPHIE; and light curves from the initial TESS Sector 15 transit, the subsequent NEOSSat detection, and also the much later Sector 56 detection by TESS.  
The parameter posterior results are listed in Table~\ref{tab:TOI2010}.
The fits converged fully by two different statistics: the Gelman--Rubin statistic, R$z$, and the number of independent samples, T$z$.  
We set very stringent thresholds of R$z<1.01$ and T$z>1000$.
We provide a brief description of the steps involved, but for precise details on the internal operations of {\tt EXOFASTv2} please consult the primary paper by \citet{Eastman:2019}.

Beyond {\tt EXOFASTv2}, we conduct several other independent analyses.
We model the bulk metallicity of planet b with a custom software, and we analyse the photometric modulation of TOI-2010 to assess the stellar rotation.
In discovering a slight acceleration across the RV measurements, we also conduct a search of mass--orbit parameter space to determine what type of additional companion could be the cause. 
%

% Noticing a slight residual acceleration in the RV data, we also explore the evidence for there being another sizeable object in the system.
% We calculate the relative likelihood of different
% mass--orbit combinations for such an additional unseen companion.

% The discrepancies in the $v\sin i_\star$ measures from our various spectral measurements prompted us to look into photometric modulation of the star.
% We carry out an independent light curve modulation analysis using TESS and WASP light curves to extract the stellar rotation rate.  

% \begin{itemize}
%     \item https://ui.adsabs.harvard.edu/abs/2019arXiv190709480E/abstract
%     \item fully converged by 2 different statistics (should name them)
% \end{itemize}

%% ------------------------------------ %%
\subsection{{\tt EXOFASTv2:} SED/MIST Stellar Modeling} \label{ssec:SED/MIST}

{\tt EXOFASTv2} fetches archival photometry from 
Galaxy Evolution Explorer \citep[GALEX;][]{Bianchi_etal_2011}, 
Tycho-2 \citep{Hog_etal_2000}, 
UCAC4 \citep{Zacharias_etal_2012}, 
APASS \citep{Henden_etal_2016}, 
the Two Micron All Sky Survey \citep[2MASS;][]{Cutri_etal_2003}, 
the Wide-field Infrared Survey Explorer \citep[WISE;][]{Cutri_etal_2013}, 
Gaia \citep{Gaia_collaboration_2016}, 
the Kepler INT Survey \citep{Greiss_etal_2012}, 
the UBV Photoelectric Catalog \citep{Mermilliod_1994}, 
and the Stroemgren--Crawford $uvby\beta$ photometry catalog \citep{Paunzen_2015}, 
as well as extinctions from \citet{Schlegel_etal_1998} and \citet{Shlafly_Finkbeiner_2011}
and parallaxes from Gaia DR2 \citep{Gaia_DR2}.
Allowing photometric uncertainties to be inflated in case of underestimation, it then fits an SED model to this archival photometry using the parallax value and a library of stellar atmospheres.  
The stellar physics are constrained from either the empirical relations laid out by \citet{Torres_etal_2010}, the Yonsie Yale stellar evolutionary models \cite{Yi_2001}, or the MIST evolutionary models 
\citep{Choi_etal_2016,Dotter:2016},
which itself is built using MESA 
\citep{Paxton_etal_2011,Paxton_etal_2013,Paxton_etal_2015,Paxton_etal_2018}.
Stellar atmospheric models from NextGen 
\citep{Allard_etal_2012}), ATLAS \citep{Kurucz_1979}, 
and PHOENIX 
\citep{Hauschildt_1997}
underlie several aspects of the code.
%

% {\tt EXOFASTv2} simultaneously derives other stellar parameters using MIST isochrones
% \citep{Dotter:2016,Choi_etal_2016,Paxton_etal_2011,Paxton_etal_2013,Paxton_etal_2015,Paxton_etal_2018}.
We are able to impose Gaussian priors on the 
stellar effective temperature ($T_{\rm eff}$)
and metallicity ([Fe/H]) in the fit, originating from the Keck/HIRES spectra (Table~\ref{tab:spec_params}). 
%
% \cmc{(Note from Jason Eastman: The log$g$ from spectra is bad and should be ignored (see \url{https://ui.adsabs.harvard.edu/abs/2012ApJ...757..161T/abstract}).
% The log$g$ from MIST+SED is better. 
% The log$g$ from transits is way better (see \url{https://arxiv.org/abs/2209.14301}).
% }
%
The $F_{\rm bol}$ and $T_{\rm eff}$ parameters have enforced error floors representative of the systematic uncertainties between stellar evolution models to prevent unrealistic precision \citep[2.0\% and 2.4\%, respectively;][]{Tayar_etal_2022}.

%
%% This gives us essentially all the stellar parameters we could want, except vsini and magnitudes (but we could list those separately if desired).
%
%% Figure to look at here (or include) is TOI2010.mcmc.mist.eps

%% ------------------------------------ %%
\subsection{{\tt EXOFASTv2:} RV Modeling}

% The RV modeling is fairly straightforward.  
The multiple instrument RV data sets are simultaneously fit to a Keplerian model, retaining separate jitter and systemic offset terms.  
We measured typical $\log R'_{HK}$ activity index measures of $-5.1$ to $-4.7$ that were uncorrelated with RV values.  This activity level could induce stellar jitter up to $\sim$10\,m\,s$^{-1}$, but is unlikely to affect the derived parameters given such a strong planetary signal.

Within {\tt EXOFASTv2} the exoplanet mass radius relation from 
\citet{Chen_Kipping_2017}
can be referenced to estimate the mass or radius of the exoplanet (and all relevant derived parameters) in the absence of an RV data set or transit, respectively. 
In this case, however, the RV data constrain the mass while the transit data constrain the radius.

A single long-term linear drift parameter is included in the model.  
We have excellent temporal overlap of the data across instruments, so there is no large correlation between the trend parameter and systemic RV offset parameters.
The fitted RV model is shown in Figure~\ref{fig:RVs}.

%% The output figure to see here is TOI2010.mcmc.rv.eps.

%% ------------------------------------ %%
\subsection{{\tt EXOFASTv2:} Transit Modeling} \label{ssec:transit_modelling}

{\tt EXOFASTv2}'s transit model is generated using
\citet{Mandel_Agol_2002} and \citet{Agol_etal_2019}
with limb-darkening parameters constrained by 
\citet{Claret_Bloemen_2011} and \citet{Claret_2017}. 
We pass it the TESS coverage of the initial Sector 15-detected transit,  a stretch of flat light curve from a Sector 40 that narrowly missed another transit, as well as the later detection in Sector 56.  We also include the entire 6\,day NEOSSat light curve with its detection.
Limb-darkening parameters and transit depths are allowed to differ between instruments.  Each instrument also gets its own jitter parameter and out-of-transit offset value.  
We impose no additional transit-specific priors for this portion of the fit.  
The period and other orbital element constraints arise from a simultaneous fit of the transit and RV data.

We incorporate a spline fit in the {\tt EXOFASTv2} modeling of the NEOSSat data, based on the {\tt keplerspline}\footnote{ \url{https://github.com/avanderburg/keplerspline}} 
\citep{Vanderburg_etal_2016}
designed to handle long-term variability in long Kepler light curves. 
We used a knot spacing of 1.1 days (roughly 3$\times$ the transit duration) to model the low-frequency variation.

% Requiring long baselines for adequate spline fitting, we include the full NEOSSat light curve.
%
% The NEOSSat data has substantially poorer SNR than the TESS detection, so we are largely relying on the TESS light curve to constrain the transit shape whereas the NEOSSat addition will provide a vastly improved period constraint over the RVs alone.
% \cmc{(Need to re-work this bit in light of new TESS detection)}
%% Improve precision by factor 500!!

%% pertinent figure here is TOI2010.mcmc.transit.eps (exofast output) or NEOSSat_transit_v1.pdf (Paul's creation)

\subsection{Bulk Planetary Composition}

To infer the bulk composition of the planet, we use the modeling and retrieval approach of \citet{Thorngren_Fortney_2019}, which we will briefly summarize.  
Forward models parameterize the thermal state of the planet by the envelope specific entropy, which we evolve from a hot initial state using the atmosphere models of \citet{Fortney_etal_2007}.  
This requires that we know the radius and temperature structure of the planet at a given specific entropy.  
We calculate this using a 1D static model of the planet which solves the equations of hydrostatic equilibrium, conservation of mass, and the equation of state (EOS).  
We use the H/He EOS from \citet{Chabrier_etal_2019}, and a 50/50 rock/ice mixture for the metals \citep{Thompson_1990}, combining these using the additive volumes approximation.   
For a given mass, metallicity, and stellar insolation, this yields evolution tracks of the radius with time.  
To match these models to TOI-2010\,b, we use a Bayesian statistical model with the true mass, bulk metallicity, and true age as model parameters and fit them against the observed mass, radius, and age 
from the EXOFASTv2 fit 
(Table~\ref{tab:TOI2010}).
Because TOI-2010\,b is much cooler than the hot-Jupiter inflation threshold \citep[e.g.][]{Miller_Fortney_2011}, we do not include any additional heating in the planet.

%% Modulation analysis
\subsection{TESS Light Curve Modulation}
Even in the PDCSAP TESS data, with a degree of its systematics removed, we noticed some low-level variability.
As a secondary measure of stellar rotation we looked at the star's long-term light curve modulation.
A simple normalization was applied in order to concatenate the TESS 2 minute cadence SAP light curves from Sectors 14, 15, and 16. Data points that were flagged as poor quality, greater than 5$\sigma$ outliers, or during the TOI-2010\,b transit were removed. 
Stitching together multiple sectors of observations (even when observed continuously) can introduce systematics into the concatenated light curve that could produce a spurious signal in a periodogram search. 
Therefore, we searched for periodic photometric variability using the TESS systematics-insensitive periodogram tool, \texttt{TESS-SIP}\footnote{\url{https://github.com/christinahedges/TESS-SIP}} \citep{Hedges_etal_2020}, which uses PCA to account for spacecraft systematics while simultaneously performing a periodogram search. 
In the periodogram search from 1 day to half the baseline of the continuous TESS observations (which are $\sim$32\,days), 
% \cmc{(This must be the half, or we wouldn't see a 19-day signal in a 16-day baseline)}
we identify a periodic signature in the light curve at 
% $18.974 \pm 2.995$\,days 
$19.0 \pm 3.0$\,days, 
albeit at a low normalized Lomb--Scargle power ($<$0.001). 
Assuming a small stellar obliquity, this 19\,day signal corresponds to a $\sim$2.8\,km\,s$^{-1}$ stellar rotation rate, in general agreement with the spectroscopic estimates (see Table~\ref{tab:spec_params}).
A short-period periodogram search (0.01--13\,days) was also performed separately on the PDCSAP photometry from Sectors 14, 15, and 16 following the procedure described in \citet{Fetherolf_etal_2023}.
A small-amplitude ($\lesssim$\,0.2 ppt), 5.7\,day signal was identified in the TESS photometry but we note this is consistent with being attributed to spacecraft systematics due to its location in power--period space relative to other stars in these TESS sectors.
With its low SNR and similarity to known systematics, we do not consider this signal physically relevant.

\subsection{WASP Light Curve Modulation}

%
%% Modulation search
We searched each season of WASP data for a rotational modulation using methods discussed in 
% \citet{2011PASP..123..547M}
\citet{Maxted_etal_2011}. We find a significant and persistent modulation at a period of 20 $\pm$ 1 days. 
% \cmc{(Fig.~\ref{fig:wasp})}. 
The modulation is weak, with an amplitude of only 1--2\,mmag, but the overall false-alarm likelihood is below 1\%.
This closely matches the TESS photometric modulation, and also likely reflects a stellar rotation rate of $\sim$2.8 km s$^{-1}$.

\begin{figure}[t]
	\centering
 % Trim:  Left  Bottom  Right   Top
	\includegraphics[width=0.47\textwidth,
    trim=35 40 100 40,clip]
    {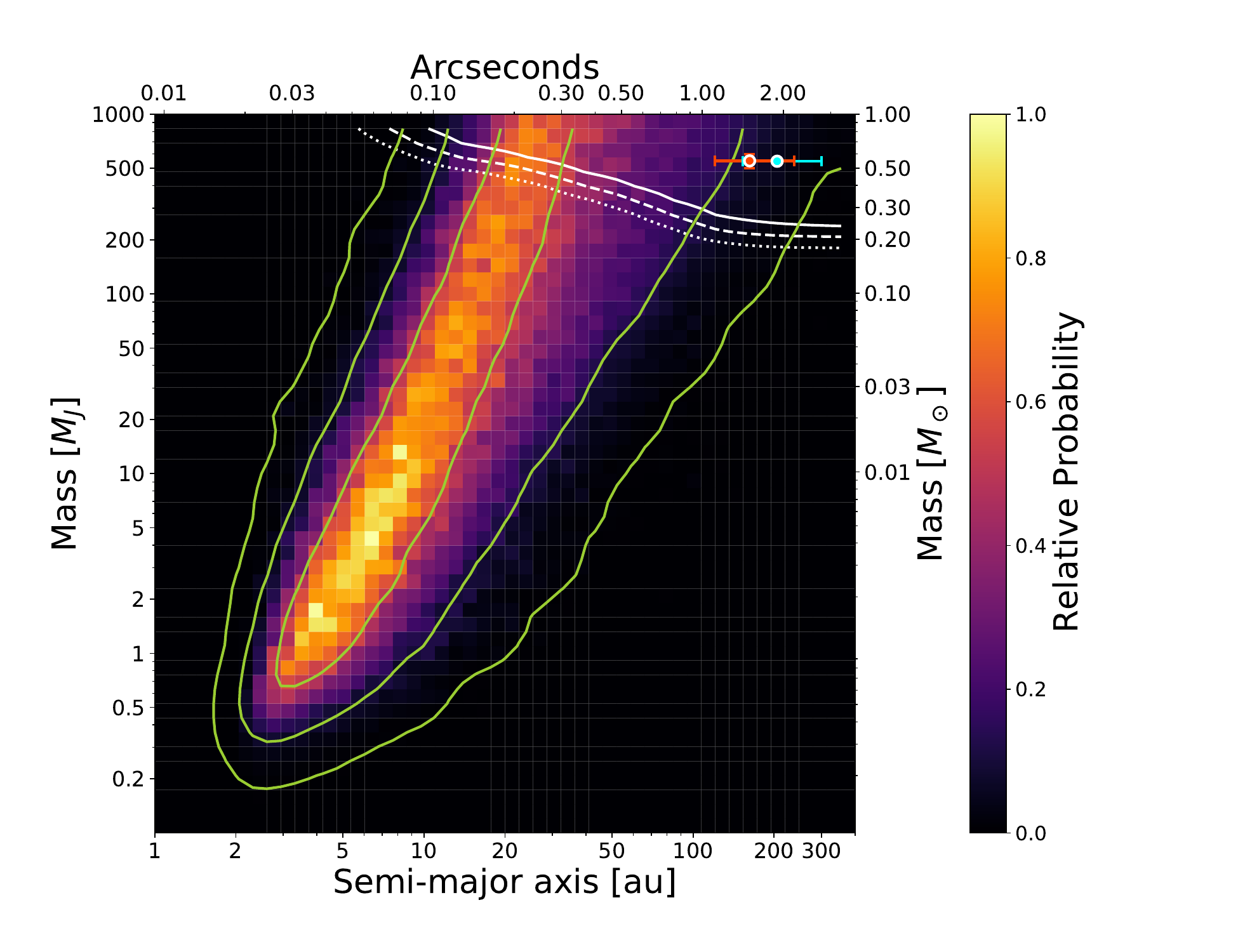}
    
    \caption{
    Constraints on a potential second bound body in the system creating the observed long-term RV acceleration, assuming the objects are at the same distance as TOI-2010.  Small-orbit limits are provided by the long baseline over which the gradual acceleration occurs.  
    Green contours show the
    0.607, 0.135, and 0.011 
    relative probability levels, corresponding to the 1, 2, and 3$\sigma$ probability density values of a normal distribution.  
    White contours show the 25\%, 50\%, and 75\% detection probability levels (from bottom to top) set by the Gemini-N/`Alopeke contrast curve.  Contours are smoothed by a Gaussian kernel with a standard deviation of one cell.
    Red and cyan points in the top right corner indicate the `Alopeke-discovered (1\farcs5) and Gaia (1\farcs9) close neighbour stars. 
        \label{fig:planet2}}
\end{figure}

\subsection{Mass--Orbit Possibilities for an Additional Companion}\label{ssec:grid-search}

A small residual acceleration is detected in the collective RV data.
To constrain potential objects on very long orbits that could cause this acceleration, we adopt the process described in \citet{Bryan_etal_2016}.
In essence, we step through a 2D grid of semi-major axes and object masses.  
In each cell, we draw a set of planet b parameters from our fitted posteriors of Table~\ref{tab:TOI2010}, generating an RV model. 
This model is subtracted from the RV measurements to reveal a residual slope.
A model for object 2 is created by drawing $M$ and $a$ values from the current cell, and $i$ and $e$ values from motivated distributions.  
In this case, $i$ is drawn randomly from a uniform $\cos i$ distribution and $e$ from a $\beta$ distribution \citep[eq. 3 of ][]{Bryan_etal_2016}.
The remaining $T_0$, $T_{\rm peri}$, and $\omega$ parameters are determined by fitting the drawn object 2 model to the residuals, as well as an RV offset.  The offset accounts for any change to the assumed systemic velocity caused by the second object.  

This process is repeated 500 times per cell.  For each cycle, the $\chi^2$ is calculated and stored, as well as the probability that the object is visible.  
This probability is based on the inferred magnitude of the object and the fraction of its orbital period that it would spend at a detectable separation from the host, set by the Gemini-N/`Alopeke contrast curve (Figure~\ref{fig:Gemini_speckle}).
%
% \cmc{The $\chi^2$ and visibility probability cubes are combined,} 
%
The data cubes are marginalized over the third axis to create a 2D probability density grid over the object's $a$ and $M$.  These results are displayed in Figure~\ref{fig:planet2}.

With the allowable mass--orbit space mapped out, we investigate the potential for the known nearby stars to cause the acceleration.
%
%% Concentration of probability in at the lower left corner is to do with eccentricity.  If we fix e=0, then the whole range has ~equal density. Fixing i=90 doesn't really change things.
%
The plotted mass uncertainties for the two nearby sources in Figure~\ref{fig:planet2} reflect the magnitude uncertainty of the source only.  
They do not account for the intrinsic uncertainty of the precalculated stellar evolution grids\footnote{\url{http://perso.ens-lyon.fr/isabelle.baraffe/BHAC15dir/}} \citep{Baraffe_etal_2015} from which they were interpolated.
%
%%% grids came from BHAC15 iso.CIT2 (http://perso.ens-lyon.fr/isabelle.baraffe/BHAC15dir/)
%%% accessed from https://species.readthedocs.io/en/latest/overview.html

To place a tentative uncertainty on the semi-major axis of the 
% Gaia and `Alopeke 
neighbouring stars (under the assumption they are bound), we follow the steps of \citet[][Appendix A2]{Brandeker_etal_2006} given an observed projected separation, unknown orbital orientation, and assumed eccentricity distribution.  
We do not adopt their analytic approximation (Equation (A2)) which roughly matches the numerical distribution arising from a simple $f(e) = e^2$ eccentricity distribution, but instead carry out the full Monte Carlo approach to create a nonanalytic distribution.  
We use $f(e) = e^{0.4}$ where $e \in [0,0.8]$ is extracted from the observation of binaries with Sun-like primaries \citep{Moe_DiStefano_2017}.

%%%%%%%%%%%%%%%%%%%%%%%%%%%%%%%%%%%%%%%%%%%%%%%%%%
%%%%%%%%%%%%%%%%%%%%%%%%%%%%%%%%%%%%%%%%%%%%%%%%%%
%%%%%%%%%%%%%%%%%%%%%%%%%%%%%%%%%%%%%%%%%%%%%%%%%%

% \documentclass{aastex62}
\providecommand{\bjdtdb}{\ensuremath{\rm {BJD_{TDB}}}}
\providecommand{\feh}{\ensuremath{\left[{\rm Fe}/{\rm H}\right]}}
\providecommand{\teff}{\ensuremath{T_{\rm eff}}}
\providecommand{\teq}{\ensuremath{T_{\rm eq}}}
\providecommand{\ecosw}{\ensuremath{e\cos{\omega_\star}}}
\providecommand{\esinw}{\ensuremath{e\sin{\omega_\star}}}
\providecommand{\msun}{\ensuremath{\,M_\Sun}}
\providecommand{\rsun}{\ensuremath{\,R_\Sun}}
\providecommand{\lsun}{\ensuremath{\,L_\Sun}}
\providecommand{\mj}{\ensuremath{\,M_{\rm J}}}
\providecommand{\rj}{\ensuremath{\,R_{\rm J}}}
\providecommand{\me}{\ensuremath{\,M_{\rm E}}}
\providecommand{\re}{\ensuremath{\,R_{\rm E}}}
\providecommand{\fave}{\langle F \rangle}
\providecommand{\fluxcgs}{10$^9$ erg s$^{-1}$ cm$^{-2}$}
% \usepackage{apjfonts}
% \begin{document}
\startlongtable
\begin{deluxetable*}{lccccc}
\tablecaption{Median Values and 68\% Confidence Interval for TOI-2010.}
\tablehead{\colhead{~~~Parameter} & \colhead{Units} & \multicolumn{4}{c}{Values}}
\startdata
%
% --- Stellar parameters ---
%
\smallskip\\\multicolumn{2}{l}{Stellar Parameters:}&\smallskip\\
~~~~$M_\star$\dotfill &Mass (\msun)\dotfill &$1.112^{+0.048}_{-0.055}$\\
~~~~$R_\star$\dotfill &Radius (\rsun)\dotfill &$1.079^{+0.027}_{-0.026}$\\
~~~~$R_{\star,SED}$\dotfill &Radius (\rsun)\dotfill &$1.0753^{+0.0093}_{-0.0090}$\\
~~~~$L_\star$\dotfill &Luminosity (\lsun)\dotfill &$1.299^{+0.083}_{-0.081}$\\
~~~~$F_{Bol}$\dotfill &Bolometric flux (cgs)\dotfill &$0.00000000354^{+0.00000000023}_{-0.00000000022}$\\
~~~~$\rho_\star$\dotfill &Density (cgs)\dotfill &$1.243^{+0.087}_{-0.086}$\\
~~~~$\log{g}$\dotfill &Surface gravity (cgs)\dotfill &$4.417^{+0.021}_{-0.025}$\\
~~~~$T_{\rm eff}$\dotfill &Effective temperature (K)\dotfill &$5929\pm74$\\
% ~~~~$T_{\rm eff,SED}$\dotfill &Effective Temperature$^{1}$ (K)\dotfill &$5950\pm110$\\
~~~~$[{\rm Fe/H}]$\dotfill &Metallicity (dex)\dotfill &$0.168\pm0.055$\\
~~~~$[{\rm Fe/H}]_{0}$\dotfill &Initial metallicity$^{1}$ \dotfill &$0.154^{+0.054}_{-0.055}$\\
~~~~Age\dotfill &Age (Gyr)\dotfill &$1.9^{+2.2}_{-1.3}$\\
% ~~~~$EEP$\dotfill &Equal Evolutionary Phase$^{3}$ \dotfill &$335^{+25}_{-37}$\\
~~~~$A_V$\dotfill &$V$-band extinction (mag)\dotfill &$0.210^{+0.079}_{-0.085}$\\
~~~~$\sigma_{SED}$\dotfill &SED photometry error scaling \dotfill &$0.72^{+0.23}_{-0.15}$\\
~~~~$\varpi$\dotfill &Parallax (mas)\dotfill &$9.237\pm0.017$\\
~~~~$d$\dotfill &Distance (pc)\dotfill &$108.26\pm0.20$\\
~~~~$\dot{\gamma}$\dotfill &RV slope$^{2}$ (m\,s$^{-1}$\,day$^{-1}$)\dotfill &$0.0185^{+0.0055}_{-0.0054}$\\
%
%
% --- Planet parameters ---
%
%
\smallskip\\\multicolumn{2}{l}{Planetary Parameters:}&b\smallskip\\
~~~~$P$\dotfill &Period (days)\dotfill &$141.834025^{+0.000065}_{-0.000066}$\\
~~~~$R_P$\dotfill &Radius (\rj)\dotfill &$1.054\pm0.027$\\
~~~~$M_P$\dotfill &Mass (\mj)\dotfill &$1.286^{+0.055}_{-0.057}$\\
~~~~$T_C$\dotfill &Time of conjunction$^{3}$ (\bjdtdb)\dotfill &$2458712.30168^{+0.00042}_{-0.00041}$\\
% ~~~~$T_T$\dotfill &Time of minimum projected separation$^{6}$ (\bjdtdb)\dotfill &$2458712.30168^{+0.00042}_{-0.00041}$\\
% ~~~~$T_0$\dotfill &Optimal conjunction Time$^{7}$ (\bjdtdb)\dotfill &$2459421.47181\pm0.00027$\\
~~~~$a$\dotfill &Semi-major axis (AU)\dotfill &$0.5516^{+0.0078}_{-0.0093}$\\
~~~~$i$\dotfill &Inclination$^{4}$ (Degrees)\dotfill &$89.903^{+0.064}_{-0.059}$\\
~~~~$e$\dotfill &Eccentricity \dotfill &$0.212^{+0.022}_{-0.021}$\\
~~~~$\omega_\star$\dotfill &Argument of periastron (Degrees)\dotfill &$98.8^{+4.8}_{-4.9}$\\
~~~~$T_{eq}$\dotfill &Equilibrium temperature$^{5}$ (K)\dotfill &$400.2^{+5.6}_{-5.7}$\\
~~~~$\tau_{\rm circ}$\dotfill &Tidal circularization timescale (Gyr)\dotfill &$3980000^{+910000}_{-790000}$\\
~~~~$K$\dotfill &RV semiamplitude (m/s)\dotfill &$47.8\pm1.5$\\
~~~~$R_P/R_\star$\dotfill &Radius of planet in stellar radii \dotfill &$0.10035^{+0.00043}_{-0.00037}$\\
~~~~$a/R_\star$\dotfill &Semi-major axis in stellar radii \dotfill &$109.8^{+2.5}_{-2.6}$\\
~~~~$\delta$\dotfill &$\left(R_P/R_\star\right)^2$ \dotfill &$0.010069^{+0.000087}_{-0.000075}$\\
~~~~$\delta_{\rm NEOSSat}$\dotfill &Transit depth in NEOSSat (fraction)\dotfill &$0.01284^{+0.00040}_{-0.00038}$\\
~~~~$\delta_{\rm TESS}$\dotfill &Transit depth in TESS (fraction)\dotfill &$0.01160\pm0.00011$\\
~~~~$\tau$\dotfill &Ingress/egress transit duration (days)\dotfill &$0.03363^{+0.0013}_{-0.00069}$\\
~~~~$T_{14}$\dotfill &Total transit duration (days)\dotfill &$0.3617^{+0.0012}_{-0.0010}$\\
~~~~$T_{FWHM}$\dotfill &FWHM transit duration (days)\dotfill &$0.32784^{+0.00083}_{-0.00082}$\\
~~~~$b$\dotfill &Transit impact parameter \dotfill &$0.147^{+0.088}_{-0.097}$\\
~~~~$b_S$\dotfill &Eclipse impact parameter \dotfill &$0.23^{+0.13}_{-0.15}$\\
~~~~$\tau_S$\dotfill &Ingress/egress eclipse duration (days)\dotfill &$0.0525^{+0.0035}_{-0.0028}$\\
~~~~$T_{S,14}$\dotfill &Total eclipse duration (days)\dotfill &$0.544^{+0.027}_{-0.025}$\\
~~~~$T_{S,FWHM}$\dotfill &FWHM eclipse duration (days)\dotfill &$0.491\pm0.025$\\
~~~~$\delta_{S,2.5\mu m}$\dotfill &Blackbody eclipse depth at 2.5$\mu$m (ppm)\dotfill &$0.0094^{+0.0020}_{-0.0017}$\\
~~~~$\delta_{S,5.0\mu m}$\dotfill &Blackbody eclipse depth at 5.0$\mu$m (ppm)\dotfill &$4.75^{+0.46}_{-0.43}$\\
~~~~$\delta_{S,7.5\mu m}$\dotfill &Blackbody eclipse depth at 7.5$\mu$m (ppm)\dotfill &$32.1^{+2.0}_{-1.9}$\\
~~~~$\rho_P$\dotfill &Density (cgs)\dotfill &$1.36^{+0.11}_{-0.10}$\\
~~~~$logg_P$\dotfill &Surface gravity \dotfill &$3.457^{+0.024}_{-0.026}$\\
% ~~~~$\Theta$\dotfill &Safronov Number \dotfill &$1.211^{+0.049}_{-0.048}$\\
~~~~$\fave$\dotfill &Incident Flux (\fluxcgs)\dotfill &$0.00557^{+0.00030}_{-0.00029}$\\
~~~~$T_P$\dotfill &Time of periastron (\bjdtdb)\dotfill &$2458572.7\pm1.2$\\
~~~~$T_S$\dotfill &Time of eclipse (\bjdtdb)\dotfill &$2458780.3^{+1.7}_{-1.6}$\\
~~~~$T_A$\dotfill &Time of ascending node (\bjdtdb)\dotfill &$2458685.2\pm1.1$\\
~~~~$T_D$\dotfill &Time of descending node (\bjdtdb)\dotfill &$2458595.5\pm1.1$\\
% ~~~~$V_c/V_e$\dotfill & \dotfill &$0.809\pm0.018$\\
~~~~$e\cos{\omega_\star}$\dotfill & \dotfill &$-0.032\pm0.018$\\
~~~~$e\sin{\omega_\star}$\dotfill & \dotfill &$0.208^{+0.022}_{-0.021}$\\
~~~~$M_P\sin i$\dotfill &Minimum mass (\mj)\dotfill &$1.286^{+0.055}_{-0.057}$\\
~~~~$M_P/M_\star$\dotfill &Mass ratio \dotfill &$0.001107^{+0.000039}_{-0.000038}$\\
~~~~$d/R_\star$\dotfill &Separation at midtransit \dotfill &$86.8\pm4.2$\\
~~~~$P_T$\dotfill &\emph{A priori} nongrazing transit prob. \dotfill &$0.01037^{+0.00053}_{-0.00048}$\\
~~~~$P_{T,G}$\dotfill &\emph{A priori} transit prob. \dotfill &$0.01268^{+0.00065}_{-0.00059}$\\
~~~~$P_S$\dotfill &\emph{A priori} nongrazing eclipse prob. \dotfill &$0.006779^{+0.00010}_{-0.000068}$\\
~~~~$P_{S,G}$\dotfill &\emph{A priori} eclipse prob. \dotfill &$0.008291^{+0.00013}_{-0.000085}$\\
%
%
% --- Wavelength parameters ---
%
%
\smallskip\\\multicolumn{2}{l}{Wavelength Parameters:}&NEOSSat&TESS\smallskip\\
~~~~$u_{1}$\dotfill &Linear limb-darkening coeff. \dotfill &$0.442\pm0.048$&$0.271\pm0.019$\\
~~~~$u_{2}$\dotfill &Quadratic limb-darkening coeff. \dotfill &$0.301\pm0.050$&$0.270\pm0.027$\\
%
%
% --- Telescope parameters ---
%
%
\smallskip\\\multicolumn{2}{l}{Telescope Parameters:}&Levy&SOPHIE&Tull\smallskip\\
~~~~$\gamma_{\rm rel}$\dotfill &Relative RV Offset$^{2}$ (m\,s$^{-1}$)\dotfill &$0.3\pm1.3$&$-15315.8^{+2.5}_{-2.4}$&$8778.8^{+2.9}_{-2.6}$\\
~~~~$\sigma_J$\dotfill &RV jitter (m\,s$^{-1}$)\dotfill &$7.81^{+1.1}_{-0.97}$&$9.9^{+2.1}_{-1.7}$&$0.00^{+8.1}_{-0.00}$\\
~~~~$\sigma_J^2$\dotfill &RV jitter Variance \dotfill &$60^{+18}_{-14}$&$97^{+46}_{-30}$&$-2^{+68}_{-39}$\\
% ~~~~\cmc{RMS}\dotfill &\cmc{Scatter around model} \dotfill &\cmc{$??^{+??}_{-??}$} &\cmc{$??^{+??}_{-??}$}&\cmc{$??^{+??}_{-??}$}\\
%
%
%
% --- Transit parameters ---
%
%
%
\smallskip\\\multicolumn{2}{l}{Transit Parameters:} & & & \\
~~TESS UT 2019-08-16 & & \\
~~~~$\sigma^{2}$\dotfill &Added variance \dotfill &$-0.000022669^{+0.000000031}_{-0.000000029}$\\
~~~~$F_0$\dotfill &Baseline flux \dotfill &$1.000132^{+0.000030}_{-0.000031}$\\
~~TESS UT 2021-07-23 & & \\
~~~~$\sigma^{2}$\dotfill &Added variance \dotfill &$-0.000000041^{+0.000000015}_{-0.000000014}$\\
~~~~$F_0$\dotfill &Baseline flux \dotfill &$1.000002\pm0.000014$\\
~~NEOSSat UT 2021-12-15 & & \\
~~~~$\sigma^{2}$\dotfill &Added variance \dotfill &$0.00001984^{+0.00000059}_{-0.00000057}$\\
~~~~$F_0$\dotfill &Baseline flux \dotfill  &$1.0003^{+0.0071}_{-0.0066}$\\
~~TESS UT 2021-09-02 & & \\
~~~~$\sigma^{2}$\dotfill &Added variance \dotfill &$0.000000077^{+0.000000060}_{-0.000000058}$\\
~~~~$F_0$\dotfill &Baseline flux \dotfill &$0.999961^{+0.000033}_{-0.000034}$\\
\enddata
\label{tab:TOI2010}
\tablenotetext{}{Notes. See Table 3 in \citet{Eastman:2019} for a detailed description of all parameters. Created using EXOFASTv2 commit number 96030ceb.
% \cmc{(Make sure note indices match the order they appear)}
}
% \tablenotetext{1}{This value ignores the systematic error and is for reference only}
\tablenotetext{1}{The metallicity of the star at birth}
% \tablenotetext{3}{Corresponds to static points in a star's evolutionary history. See \S2 in \citet{Dotter:2016}.}
\tablenotetext{2}{Reference epoch = 2459382.832843}
\tablenotetext{3}{Time of conjunction is commonly reported as the ``transit time"}
% \tablenotetext{6}{Time of minimum projected separation is a more correct ``transit time"}
% \tablenotetext{7}{Optimal time of conjunction minimizes the covariance between $T_C$ and Period}
\tablenotetext{4}{Inclination symmetrically on the other side of 90$^\circ$ is equally valid}
\tablenotetext{5}{Assumes no albedo and perfect redistribution. Calculated at a star--planet separation of $a$.  See Table~\ref{tab:orbit} for phase-specific values.}
\end{deluxetable*}
% \end{document}

%%%%%%%%%%%%%%%%%%%%%%%%%%%%%%%%%%%%%%%%%%%%%%%%%%
%%%%%%%%%%%%%%%%%%%%%%%%%%%%%%%%%%%%%%%%%%%%%%%%%%
%%%%%%%%%%%%%%%%%%%%%%%%%%%%%%%%%%%%%%%%%%%%%%%%%%

%%%%%%%%%%%%%%%%%%%%%%%%%%%%%%%%%%%%%%%%%%%%%%%%%%
%%%%%%%%%%%%%%%%%%%%%%%%%%%%%%%%%%%%%%%%%%%%%%%%%%
%%%%%%%%%%%%%%%%%%%%%%%%%%%%%%%%%%%%%%%%%%%%%%%%%%

\section{Results}\label{Sec:Results}

% \cmc{
% New results section would contain the following (could be paragraphs instead of distinct sections):
% \begin{itemize}
%     \item Summary of planet parameters; addressing phase-related quantities (Table 3); eccentricity context
%     \item no bimodality in EXOFAST fit between stellar mass and age
%     \item Summary of stellar characterization (seems mostly okay as-is)
%     \item photometric modulation (TESS + WASP)
%     \item Brief summary of stellar environment (Gaia and `Alopeke neighbours).  Tiny impact `Alopeke blend causes on radius measure.
%     \item brief mention that spectral data (SOPHIE and Keck template) show no indication of stellar blends.
%     \item bulk metallicity
%     \item General conclusion (currently written at the end of planet parameter subsection)
%     \item Summarize results for companion search (RV acceleration, RV periodogram search, lack of BLS findings).
%     \item results from grid-search
% \end{itemize}
% }

%%% Planet parameters
Based on the results of the global transit, RV, and SED fit, we confirm TOI-2010\,b as a temperate Jovian exoplanet around a Sun-like star.  
We find TOI-2010\,b to have a mass of
$M_P=1.286^{+0.055}_{-0.057}\ M_{\rm J}$
and a radius of 
$R_P=1.054\pm0.027\ R_{\rm J}$.
It orbits with a period of 
$P=141.834025^{+0.000065}_{-0.000066}$ days
and an eccentricity of
$e=0.212^{+0.022}_{-0.021}$.
A full list of the fitted and calculated parameters and their uncertainties are displayed in Table~\ref{tab:TOI2010}.
We determine that the contribution to the uncertainty of $M_P$ is almost evenly split between the uncertainties on $K$ and $M_\star$, and that the uncertainties of $i$, $e$, and $P$ have a negligible impact.

The reported $T_{\rm eq}$ of Table~\ref{tab:TOI2010} comes from an assessment using a single representative star--planet distance.  
However, with nonnegligible eccentricity, several calculated parameters, $T_{\rm eq}$ included, are subject to variation with orbital phase. 
Table~\ref{tab:orbit} displays certain of these parameters at four key points in the planet's orbit.

TOI-2010\,b's moderate eccentricity falls in the $\sim$$75^{\rm th}$ percentile of giant planets with well-known masses and radii.  It is distinctly above the cluster of planets with very low eccentricities, but not so high as to stand out among the population of high-eccentricity planets.

%%%% STELLAR CHARACTERIZATION

Between the SED/MIST model fitting within {\tt EXOFASTv2} and the spectral analysis from various instruments, we determine many of TOI-2010's stellar parameters.
In the global {\tt EXOFASTv2} posteriors, we see no sign of the bimodality commonly seen between stellar mass and age.  This can arise when a star is slightly evolved and the MIST stellar evolution models experience some degeneracy near the subgiant branch \citep[e.g.][]{Dalba_GOTEM_II}.

%% Use Keck as spectral basis
% In the course of this study, we used the Keck/HIRES stellar values in the global fit as this spectral measurement has the highest resolution and SNR.
%
%% Values from Keck
From the Keck/HIRES spectral measurement, we find the stellar radius, mass, and effective temperature to be
$R_\star = 1.084^{+0.028}_{-0.027}\,R_\odot$,
$M_\star = 1.107^{+0.050}_{-0.057}\,M_\odot$, and
$T_{\rm eff} = 5917\pm75\,K$, respectively.
TOI-2010 has a surface gravity of 
$\log g = 4.412^{+0.023}_{-0.026}$ 
and a metallicity of
$[{\rm Fe/H}] = 0.169^{+0.055}_{-0.056}$.
We used these Keck/HIRES stellar values in the global fit as this spectral measurement has the highest resolution and SNR.
%
%% Disagreement of vsini between instruments
The stellar parameters determined via spectral fitting of the Keck/HIRES, LCOGT/NRES, and FLWO/TRES data sets (Table~\ref{tab:spec_params}) show close agreement in general.
However, the instruments report distinct values for the stellar rotation ($v\sin i_\star$), though all indicate a slowly rotating star ($<$5\,km\,s$^{-1}$).
The discrepancies between the instruments may be due to differences in spectral resolution and SNR, or potentially the slow rotation of the star itself.
%%% the TRES spectrum pipeline does not include photospheric turbulence in its line modeling.
For slowly rotating stars, the effects of rotationally induced line broadening can be of similar magnitude to other broadening mechanisms (e.g., thermal, pressure, and turbulence).  
Disentangling them becomes challenging, and thus $v\sin i_\star$ may be inflated if some mechanisms are not properly considered.
We therefore report the spectrally derived rotation rates as rough upper limits (Table~\ref{tab:spec_params}).

%% TESS and WASP rotational modulations
We note that the photometric modulations seen in the TESS and WASP light curves both suggest an equatorial rotation of $\sim$2.8\,km\,s$^{-1}$, falling in the middle of the spectral $v\sin i_\star$ values.  
The precise value is not central to any key findings of our study, but it does have some bearing on estimates of a potential RM signal for future endeavours.
We adopt the modulation-derived value for RM calculations in Section~\ref{ssec:future_obs} as it bypasses the line-broadening issues of a slow rotator.

%%%%%%%%%%%%%%%%%%%%%%%%%%%%%%%%%%%%%%%%%%%%%%%%%%
% \startlongtable
\begin{deluxetable}{lccccc}
\tabletypesize{\scriptsize}
\tablecaption{Orbital Information \label{tab:orbit}}
\tablehead{
  \colhead{Parameter} & 
  \colhead{Periastron} &
  \colhead{Apoastron} &
  \colhead{Transit}&
  \colhead{Eclipse}&
  \colhead{Units}}
  \startdata
Phase          & 0.016 & 0.516 & 0.000 & 0.479 & -- \\
Orbital distance  & 0.44  & 0.70  & 0.44  & 0.67  & au \\
Insolation     & 6.9   & 2.9   & 6.8   & 2.9   & $S_\oplus$ \\
$T_{\rm eq}$   & 450   & 363   & 450   & 364   & K \\
\enddata
% \tablenotetext{}{Note: uncertainties in these values are similar to }
\tablenotetext{}{Note: The $T_{\rm eq}$ calculation assumes no albedo and perfect redistribution.}
\end{deluxetable}

%%%%%%%%%%%%%%%%%%%%%%%%%%%%%%%%%%%%%%%%%%%%%%%%%%

\begin{figure}[t]
	\centering
	\includegraphics[width=0.47\textwidth,
    trim=20 10 10 0,clip]
    {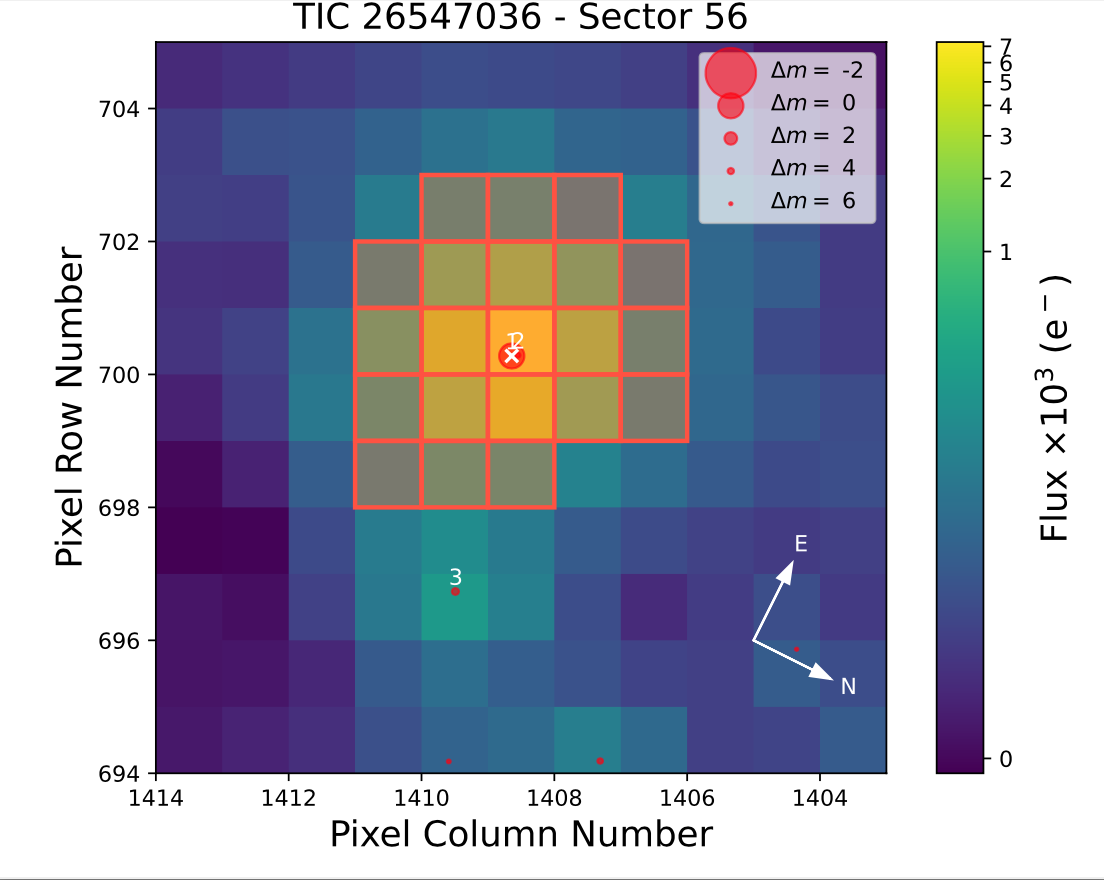}
    % Trim:  Left  Bottom  Right   Top
	
    \caption{
    Visual comparison of TESS photometry and nearby Gaia stars.  Only one notable Gaia star contaminates the aperture (labeled ``2") with a separation of 1\farcs9 and fainter by $\Delta m \sim 5$. The TESS PDCSAP flux values account for this very minor dilution.
        \label{fig:TESS_aper}}
\end{figure}

%% Stellar environment (neighbour stars)
Available data on the local stellar environment reveals two faint neighbours to TOI-2010.
We use {\tt tpfplotter}%
\footnote{\url{https://github.com/jlillo/tpfplotter}}
\citep{tpfplotter} 
to  jointly visualize the TESS aperture and Gaia positional information (Figure~\ref{fig:TESS_aper}). 
The only Gaia star of note within the TESS aperture is 5\,mag fainter and separated from TOI-2010 by 1\farcs9 at a position angle of 33$^\circ$ east of North (Gaia ID: 2136815881247621760).
The PDCSAP flux used in our light curve analysis accounts for this minor dilution so as not to affect the radius estimate of the planet.

Though absent from the Gaia catalogue, we photometrically detect a second neighbour in the immediate vicinity of TOI-2010 using `Alopeke high-contrast imaging.
It is of similar brightness to the Gaia star ($\Delta {\rm mag} = 5 \pm 0.5$) and was detected at a separation of 1\farcs5, 138$^\circ$ east of North.
This star does not appear in the Gaia catalogue.
The flux dilution from the `Alopeke star is small enough to cause $<1\%$ deviation in the $R_p/R_\star$ measurement.

%% Further unresolved neighbours via spectra
Spectral investigation for evidence of blended binaries also shows no indication of significant contamination.
%
% \cmc{(This has analysis, but not presented elsewhere)}
In the SOPHIE spectra, the corresponding bisectors of the cross-correlation functions do not show any significant variation nor correlation with the RV.
This means there are no indications for RV variations induced by blend configurations or stellar activity. 
We also computed cross-correlations using masks characteristic of different spectral types: all produce similar RV variations, suggesting against the presence of a blend of stars with different spectral types. 
%
% \cmc{(Again, this has some analysis, but not presented elsewhere)}
Similarly, the Keck template spectrum was run through the {\tt ReaMatch} \citep{ReaMatch2015} software to check for blended stellar spectra. The analysis revealed no hint of any such blended components in its cross-correlation routines, and limits any unresolved sources to well below 1\% of TOI-2010's flux.

With the lack of any photometric or spectroscopic evidence for significant or problematic nearby or blended stars, we conclude that the measurements of TOI-2010 are free of any significant stellar contamination.  The planet radius assessment of TOI-2010\,b is therefore robust.

\begin{figure}[t]
	\centering
	\includegraphics[width=0.47\textwidth]
    {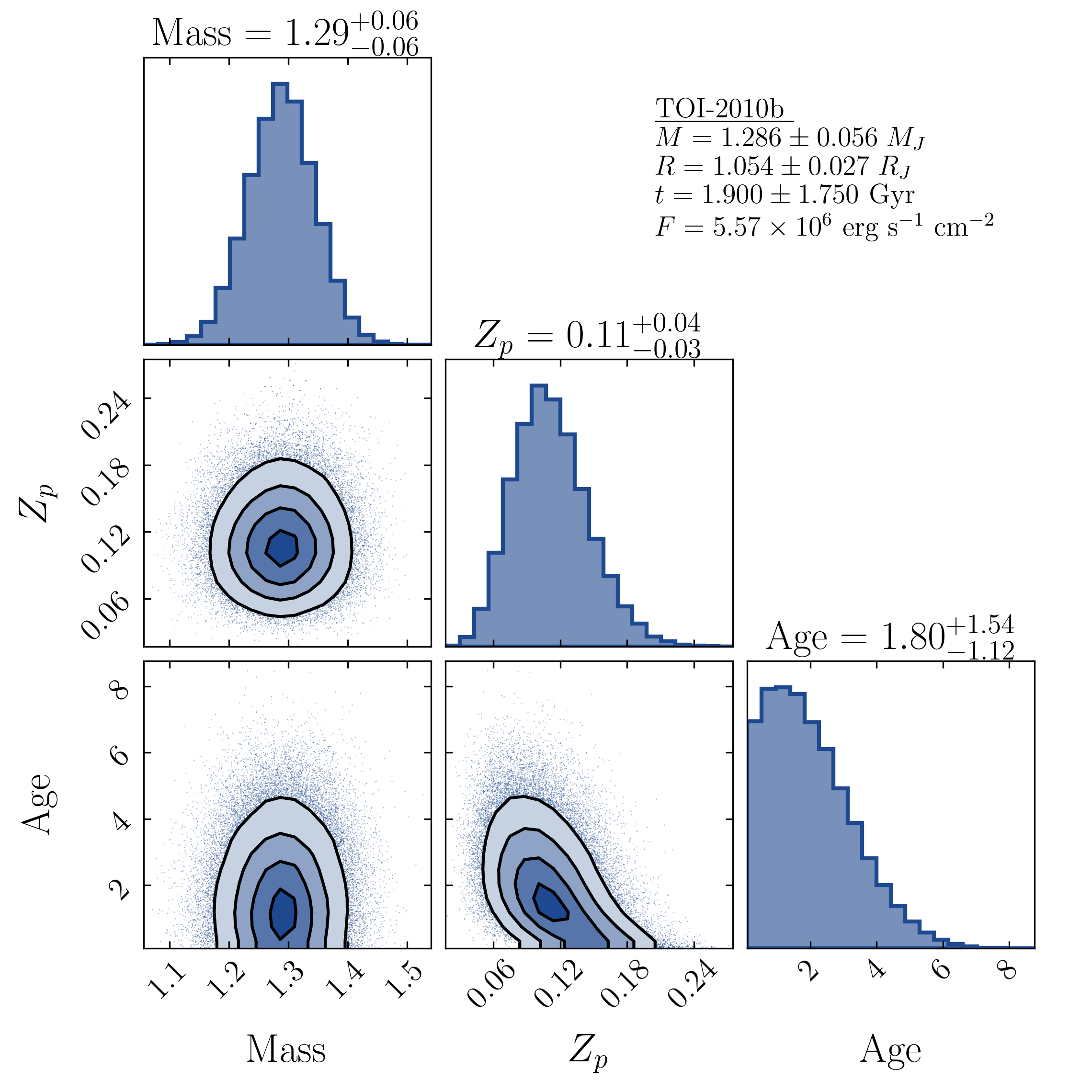}
	
    \caption{
    Results of the planet's bulk metallicity analysis.  Here mass is reported in Jupiter units, $Z_P$ is the bulk metal mass fraction of the planet, and the age is given in gigayears.  The small inset table shows the input priors used.
        \label{fig:Daniel_model}}
\end{figure}

%%% Bulk metallicity (Daniel's results)
The bulk metallicity results are shown in Figure~\ref{fig:Daniel_model}.  
The bulk metal mass fraction of $Z_P = 0.11$ corresponds to 45 Earth masses of heavy elements.
There appears to be a small degeneracy between age and metallicity in this analysis.  
This arises because leftover heat from formation in a young star ($<1$\,Gyr) is compensated in the model with extra metal.

In this case we used a fully mixed planet model.
Using a ``core+envelope" model would mean replacing compressible gas in the core with less compressible metals, requiring more metals to achieve the same radius.
This model would require $\sim$20\% (or $\sim$1$\sigma$) extra, according to \citet{Thorngren_etal_2016}.  
The result is similar when using a moderate number of layers making up a semiconvective staircase ``core," but when considering thousands of layers 
\citep[e.g.,][]{Lecont_chabrier_2012}
% (e.g. Leconte \& Chabrier 2013) 
cooling slows down and even more metal is required. However, simulations suggest that small layers merge quickly as the planet evolves 
\citep{Moll_etal_2017,Vazan_etal_2018},
% (Vazan 2018, Moll 2017), 
so we would not assume such an extreme case without more evidence.

%%% Search for other planets

% - RV acceleration
% - Lomb-scargle RV results
% - BLS light curve results
% - Grid-search results
%     - Limits on allowable space / visibility
%     - Inferred values for neighbour stars

\begin{figure}[t]
	\centering
	\includegraphics[width=0.47\textwidth]
    {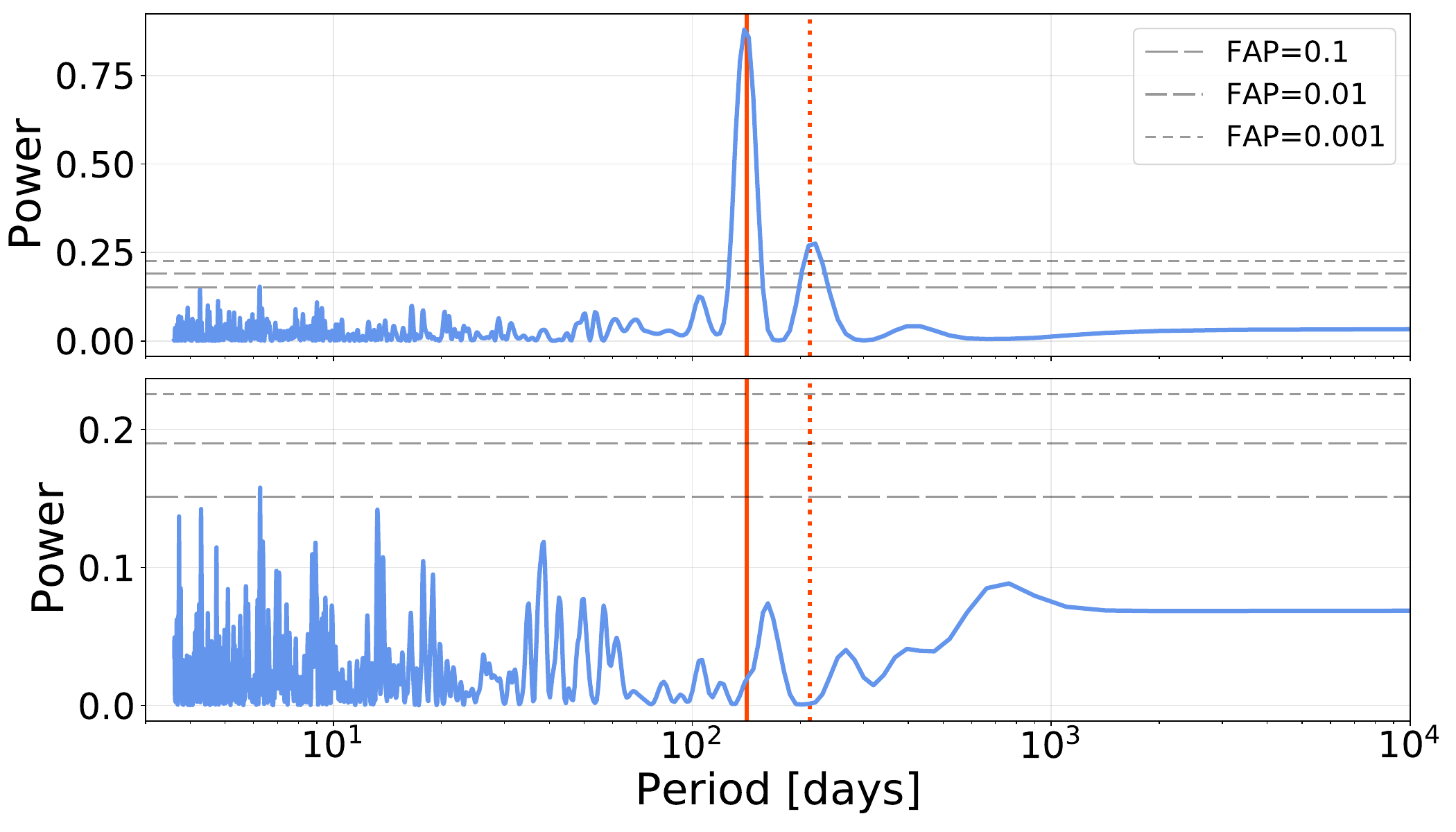}
    \caption{
    Lomb--Scargle periodogram of RV data. False-alarm probabilities (FAPs) are shown in grey.  The red solid line indicates the fitted period of TOI-2010\,b, and the dotted red line is the 3/2 harmonic.  
    The top panel periodogram shows results of the unaltered data set, whereas data used for the bottom panel has had the 141.8\,day signal removed.  With the removal of planet b's signal, no other significant power remains at any searchable period.
        \label{fig:LombScargle}}
\end{figure}

%%% RV acceleration
The RV fit reveals a residual acceleration of 
$\dot{\gamma} = 0.0185 \pm 0.0055$\,m\,s$^{-1}$\,day$^{-1}$.  
This $\sim$$3\sigma$ slope detection is 
suggestive of some additional distant planetary or stellar companion acting on the system.  
%
%%% Periodogram search
In running the RV data through a generalized Lomb--Scargle periodogram \citep[][]{Lomb_1976,Scargle_1982,Zechmeister_Kurster_2009} within the {\tt astropy} package
\citep{astropy:2013,astropy:2018,astropy:2022}, 
the obvious 142\,day signal stands out with indisputable significance.
Removing the best-fit Keplerian RV model of planet b (Figure~\ref{fig:RVs}) and rerunning the periodogram on the residuals produces no peaks with false-alarm probabilities better than $\sim$10\%.  These findings are shown in Figure~\ref{fig:LombScargle}.  
There appear to be no other periodic signals present in our data set for $P \lesssim 1000$ days. 
%

%% BLS search
We also check the TESS light curves for additional transits.
Using the {\tt astropy.stats.BoxLeastSquares} (BLS) function, we scan the available photometry for periodic transit-like signals.
With two known TOI-2010\,b transits in the data, the procedure flags the 1134\,day separation along with accompanying aliases (including the 142\,day true period).
Removing the two known transits leaves a very flat light curve, and a second BLS pass detects nothing above the noise level.
The BLS algorithm is sensitive only to repeated events, and so single transits (e.g., due to very long periods or transits falling in observing gaps) would not be detected here.
A visual inspection of the light curve reveals no obvious transits to indicate additional bodies in the system.

%%% Grid-search 
To explore the possible very-long-period scenarios, 
we employed a 2D grid search of semi-major axis and companion mass combinations that might produce the observed RV acceleration \citep[][]{Bryan_etal_2016}.
The resulting relative probability map marks correlated boundaries on the probable configurations (Figure~\ref{fig:planet2}).
%
% We find that the RVs and contrast curve allow for a correlated distribution of mass--orbit combinations.
At the low-mass end, the system could harbour an object of 0.4\,$M_{\rm J}$ orbiting at 2.6\,au. 
Anything interior and/or less massive than that struggles to match the observed acceleration.
For sources remaining below the photometric detection threshold, the most probable configuration at the high-mass end is a 475\,$M_{\rm J}$ (0.46\,$M_\odot$) object orbiting at $\sim$22\,au, however the allowable configuration space becomes quite broad.
As we are now in the range of self-luminous low-mass main-sequence stars, the Gemini-N/`Alopeke speckle imaging would likely detect anything more massive or more separated (i.e. above the white contours).

%%% Determination of neigbour star properties
In determining the plausibility of each close neighbour star causing the RV acceleration, 
we estimate some of their relevant properties.
We determine the 1\farcs5 `Alopeke star to have a mass of 
$M = 0.525^{+0.045}_{-0.049}\,M_\odot$ 
and a semi-major axis of
$a = 162.16^{+75.33}_{-41.43}$\,au, while the 1\farcs9 Gaia star has
$M = 0.5226^{+0.0001}_{-0.0001}\,M_\odot$
and 
$a = 205.21^{+95.33}_{-52.43}$\,au.
These estimates, though rough,  allow us to place both stars on the Figure~\ref{fig:planet2} grid.
% noticing that they fall in a low probability region of the explored mass--orbit space.
%

If neither of these nearby stars are the cause, the most likely candidate is a yet-unseen planetary or low-mass stellar object along the high-probability region in Figure~\ref{fig:planet2}.
This region through the explored parameter space roughly follows the trend of 
$M/[M_J] \approx 0.015\,(a/{\rm [au]})^{3.2}$
where $a > 3$\,au.

%%%%%%%%%%%%%%%%%%%%%%%%%%%%%%%%%%%%%%%%%%%%%%%%%%
%%%%%%%%%%%%%%%%%%%%%%%%%%%%%%%%%%%%%%%%%%%%%%%%%%
%%%%%%%%%%%%%%%%%%%%%%%%%%%%%%%%%%%%%%%%%%%%%%%%%%

%% --------------------------------------------------- %%
\section{Discussion}\label{Sec:Discussion}

\subsection{TOI-2010\,b in Context}

\begin{figure}
	\centering
	% Trim:  Left  Bottom  Right   Top
	\includegraphics[width=0.47\textwidth,
	trim=10 30 10 40,clip]
    {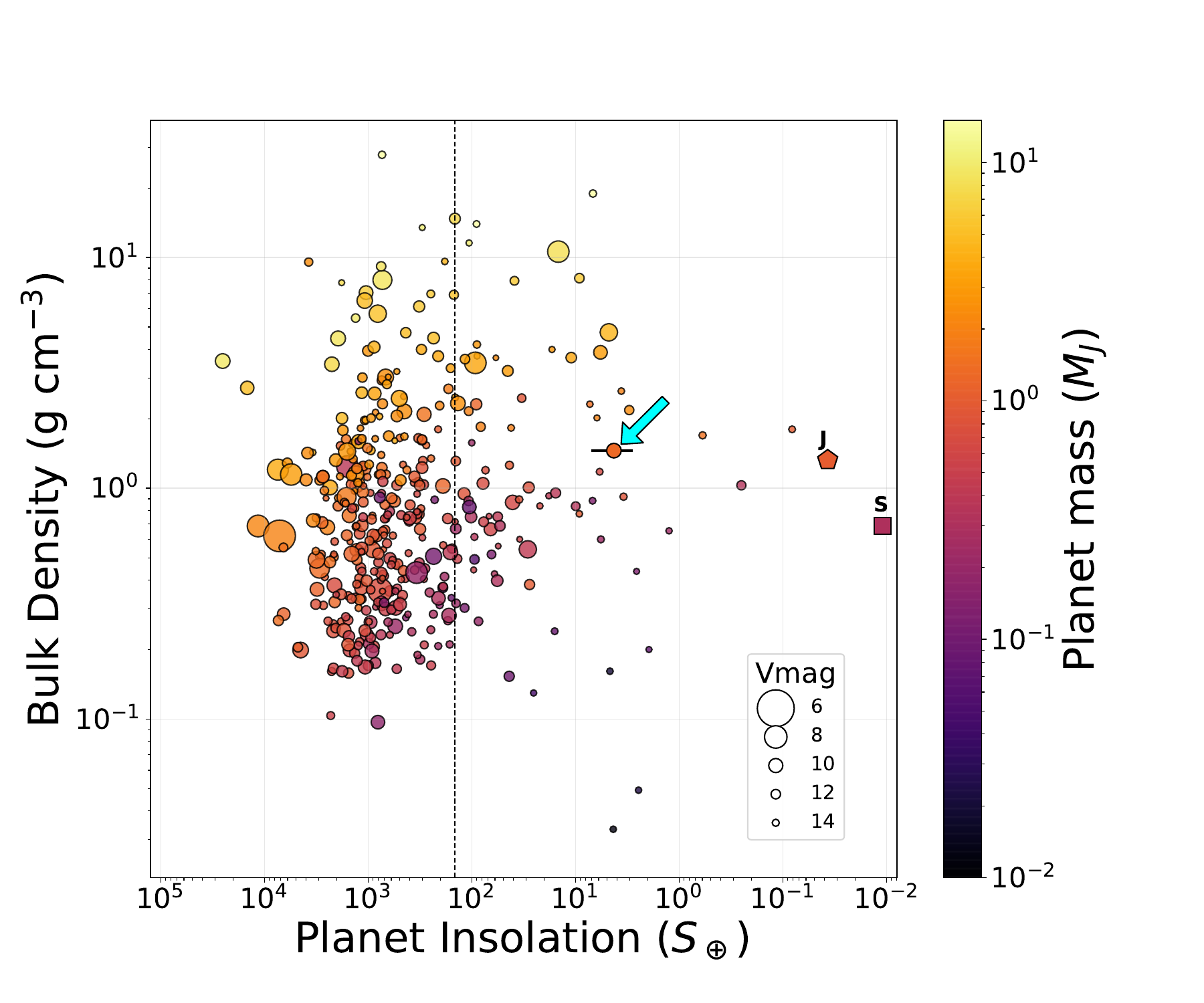}
	\includegraphics[width=0.47\textwidth,
	trim=10 25 10 40,clip]
    {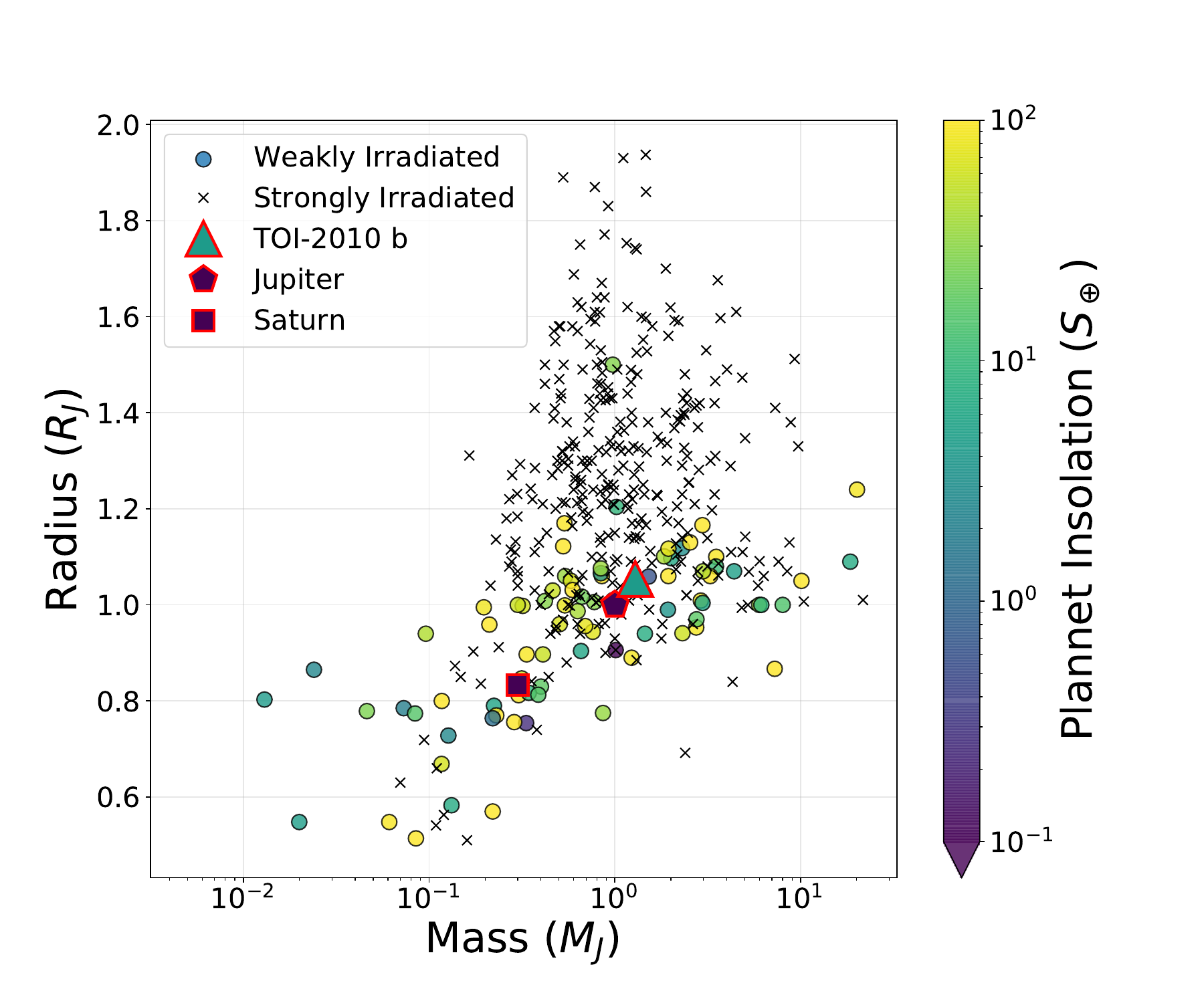}
	
    \caption{
    Population of confirmed giant ($R>0.5 R_J$) transiting planets with available insolation values and better than 50\% uncertainties on mass and radius.
    \emph{Top:} The dotted vertical line indicates the empirical inflation boundary \citep{Miller_Fortney_2011,Demory_Seager_2011} where planet radii are seen to increase with insolation.  TOI-2010\,b is indicated with the cyan arrow, and Jupiter and Saturn are labeled towards the right.  
    \emph{Bottom:} The same population of giant planets separated into strongly and weakly irradiated subgroups according to the boundary in the above plot.
        \label{fig:target_context}}
\end{figure}

With physical and orbital parameters of the TOI-2010\,b system properly constrained, we can place it in the context of other known exoplanets.  
Using planet data gathered using {\tt ExoFile}%
\footnote{\url{https://github.com/AntoineDarveau/exofile}} 
from the 
% NASA Exoplanet Archive 
\citet{NEXA_Planetary_Systems_Table},
Figure~\ref{fig:target_context} shows several properties for the population of confirmed giant planets ($R>0.5\,R_{J}$).
As can be seen in the top panel, TOI-2010\,b is deep in the low-insolation wings of the population.  Few other confirmed giants can boast such low stellar input, and fewer still have magnitudes bright enough to enable detailed spectroscopic follow up.
TOI-2010\,b stands out as valuable addition to this corner of parameter space.

The bottom panel of Figure~\ref{fig:target_context} locates TOI-2010\,b in mass and radius space.
A distinction is made between strongly and weakly irradiated planets (dotted line in the top panel and marker type in the bottom panel), given that their mass--radius relationship changes.
We find TOI-2010\,b to be a fairly typically proportioned giant planet, akin to Jupiter though quite a bit warmer.  Its moderate eccentricity may suggest a dynamic history either in its formation or due to ongoing interaction with unseen neighbours.

\subsection{Future observation potential}\label{ssec:future_obs}

\begin{figure}
	\centering
    \includegraphics[width=0.49\textwidth]
    {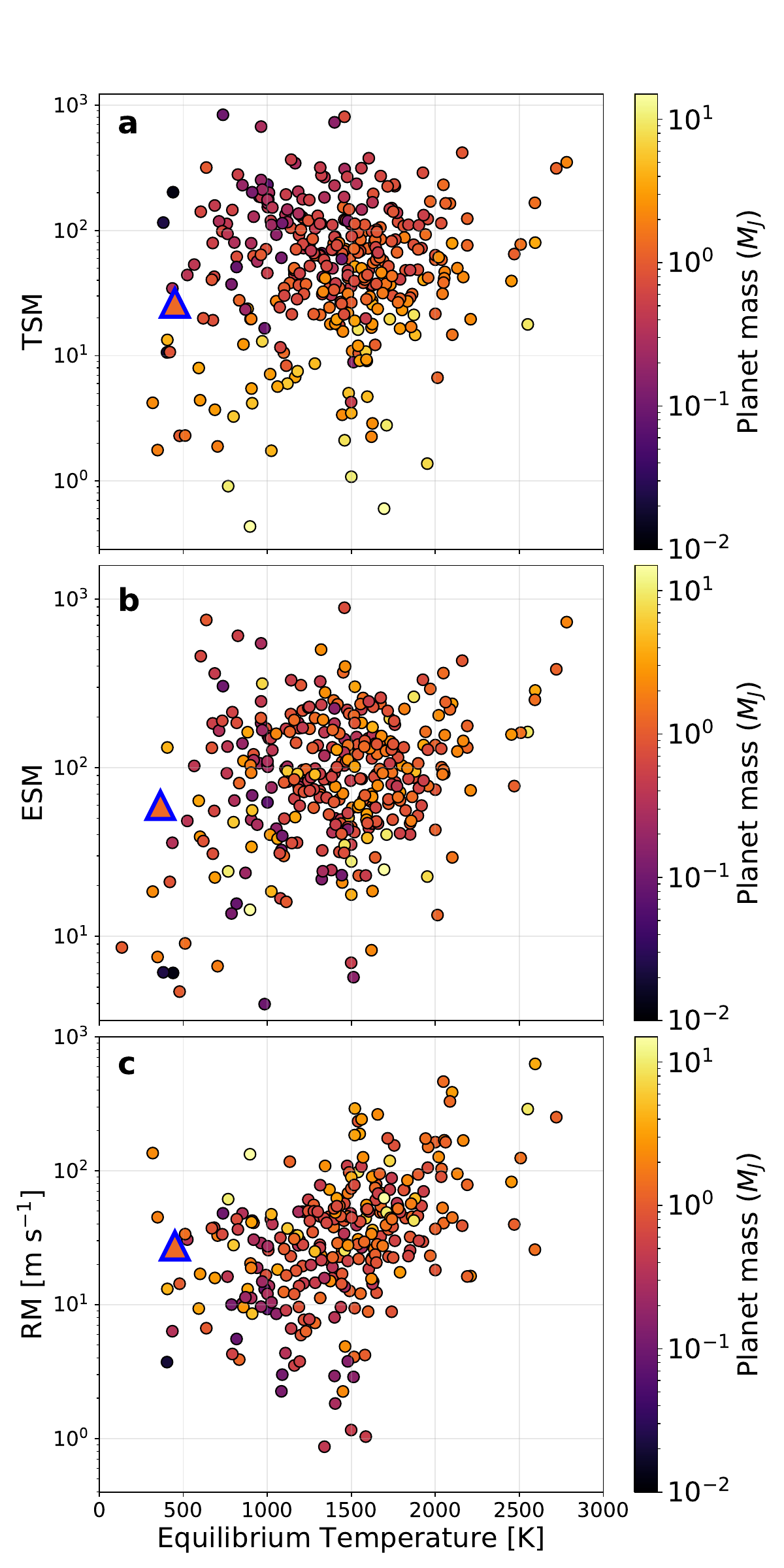}

    \caption{
    Comparison of expected signal strengths for the same planet population shown in Figure~\ref{fig:target_context}. 
    TOI-2010\,b is indicated with the triangle marker. 
    \emph{Panel a:} TSM.  
    The TSM provides an SNR estimate for a JWST/NIRISS transit observation.
    \emph{Panel b:} ESM.
    The ESM provides an SNR estimate for a JWST/MIRI eclipse observation.
    \emph{Panel c:} RM signal amplitude.  The TOI-2010\,b value in this plot uses $v\sin i_\star=2.8$ km s$^{-1}$, derived from the TESS and WASP light curve modulations.
        \label{fig:signal_strength}}
\end{figure}

We determine the expected signal strengths for a number of potential observations that might be made on this target in the future (Figure~\ref{fig:signal_strength}).
%
%% TSM
%
% Using the transit duration actually hurts the TSM
%
%%% Modified TSM - scaled by sqrt(T14) 
% The often used Transmission Spectroscopy Metric (TSM) of \citet{Kempton_etal_2018} normally provides an SNR estimate for fixed-duration observations made with JWST/NIRISS, calibrated for small planets.  
% To make meaningful comparisons of TOI-2010\,b with other giant planets, we report a modified TSM that does not produce a calibrated SNR value, but reports the relative signal strength amongst this population. 
% In addition to limiting the comparison to giant planets, we also scale each TSM value by the square-root of the planet's transit duration. This captures the observability of a planet given a single transit, rather than a fixed amount of on-target telescope time.
%
%
%%% Regular TSM (with scale-factor = 1)
We calculate and report the \citet{Kempton_etal_2018} Transmission Spectroscopy Metric (TSM), though we note it was calibrated for smaller planets.  The TSM nominally provides an SNR estimate for fixed-duration observations made with JWST/NIRISS.  However, without the small-planet-calibrated scaling factor, the specific values of Figure~\ref{fig:signal_strength}a may be better interpreted by their relative strengths, rather than absolute value.
We use TOI-2010~b's transit-phase equilibrium temperature (450 K) for this calculation, and in doing so we find that it has moderate transmission spectroscopy potential with TSM $\sim 26$.  
As a relative measure, it lands at the $\sim$23$^{\rm rd}$ percentile for the population plotted in Figures~\ref{fig:target_context} and \ref{fig:signal_strength}.

%% ESM
We can also look at the Emission Spectroscopy Metric (ESM) of \citet{Kempton_etal_2018}. 
Similarly to the TSM, the ESM estimates the SNR achieved with a mid-infrared secondary eclipse detection by JWST.
We find a more promising scenario in emission than with transmission. 
Using the eclipse-phase equilibrium temperature (364\,K), TOI-2010~b has one of the strongest predicted emission signals 
(ESM $\sim 60$) among cool giants $\lesssim 750$\,K.  Even against giant planets as a whole, TOI-2010\,b falls near the median value.  
The ESM does not include an empirical calibration like the TSM, and so the values indicate the expected SNR of a JWST secondary eclipse detection with the MIRI instrument.
TOI-2010\,b may provide a very interesting test bed for certain atmospheric properties.  The $T_{\rm eq}$ range of 360--450\,K across its orbit spans a transition regime where disequilibrium chemistry may be evident.  Models by \citet{Fortney_etal_2020} predict this temperature range to exhibit marked changes in CO/CH$_4$ and N$_2$/NH$_3$ ratios between equilibrium and disequilibrium conditions.
Such detections may go a long way toward connecting models across the Jupiter--exoplanet--brown dwarf continuum.
% Fortney et al Fig 8
%  CO/CH4=1 transition (530 K, Fig 16)
% N2/NH3 transitions ~480 K (Fig 14)

%%% Paul Dalba: Given that there is some promising outcome for future atmospheric characterization, maybe consider calling back to that Fortney et al. 2020 paper. 2010b's Teq is right in the perfect range for identification of disequilibrium CO/CH4, which directly ties into atmospheric transport. This would be a really interesting detection and would go a long way toward connecting Jupiter--exoplanets--brown dwarfs.

%% Rossiter McLaughlin
We also compute the potential for making an obliquity measurement using the RM effect.  \citet[][Equation (40)]{Winn_2010} provides an approximation for calculating the $\Delta$RV amplitude expected during transit for given planet/star size ratio, impact parameter, and projected equatorial velocity ($v\sin i_\star$) of the star.  
Adopting $v\sin i_\star = 2.8$\,km\,s$^{-1}$, derived from the modulations observed in both the TESS and WASP light curves,
% Though this value assumes a near-90$^\circ$ stellar inclination, it falls in the midst of the spread of spectroscopically-determined values (Table~\ref{tab:spec_params}).  
% We adopt this rotation estimate because it avoids the pitfalls of confounding line-broadening factors inherent in slow-rotators.
we calculate an RM amplitude of 27.5\,m\,s$^{-1}$
(Figure~\ref{fig:signal_strength}(c)).

%%%%%%%%%%%%%%%%%%%%%%%%%%%%%%%%%%%%%%%%%%%%%%%%

% \documentclass{aastex62}
% \begin{document}
% \startlongtable
\begin{deluxetable*}{lccccc}
\tablecaption{Median values and 68\% confidence interval for transit times and depths\newline }
\tablehead{\colhead{Transit} & \colhead{Planet} & \colhead{Epoch} & \colhead{$T_T$} & \colhead{Depth}}
\startdata
TESS UT 2019-08-16 & b & 0 & $2458712.30168^{+0.00042}_{-0.00041}$ & $0.011598 \pm 0.000052$ \\ % & $0.10035^{+0.00043}_{-0.00037}$\\
NEOSSat UT 2021-12-11 & b & 6 & $2459563.30584^{+0.00028}_{-0.00027}$ &  $0.01246 \pm 0.00026$ 
 \\ %& $0.10035^{+0.00043}_{-0.00037}$\\
TESS UT 2022-09-02 & b & 8 & $2459846.97389 \pm 0.00034$ &  $0.011598 \pm 0.000052$ \\% & $0.10035^{+0.00043}_{-0.00037}$\\
%
% (old)
% TESS UT 2019-08-16 & b & 0 & $2458712.30171 \pm 0.00043$ & $0.139^{+0.10}_{-0.094}$ & $0.011495 \pm 0.000076$\\
% NEOSSat UT 2021-12-15 & b & 6 & $2459564.2854^{+0.0035}_{-0.0031}$ & $0.139^{+0.10}_{-0.094}$ & $0.01240 \pm 0.00024$\\
\enddata
\label{tab:transitpars}
\end{deluxetable*}
% \end{document}

%%%%%%%%%%%%%%%%%%%%%%%%%%%%%%%%%%%%%%%%%%%%%%%%

\subsection{Single-transit Planets}

Any transit survey mission is eventually bound to produce single-transit targets.  
It comes as a direct consequence of having only finite monitoring time for a particular star, and the potential for exoplanets to have very long orbits.  
With TESS's month-long baseline coverage for most of the sky, single-transit targets are not infrequent. 
Passively detected retransits in subsequent sectors can help narrow down options, but long gaps between detections leave many possible period aliases.
The survey's schedule of reobserving a target is also not always compatible with the planet's orbit.
For example, the second detected transit of TOI-2010\,b in Sector 56 was very nearly missed.  
If the period had turned out even 0.6\% longer ($\sim$19\,hr), the additional TESS transit would have been missed entirely.

Without substantial active follow up effort, single-transit planets often remain unviable for further study.
Any phase/timing-related endeavours cannot be scheduled without a firm ephemeris, and attempting an RV study of the system requires careful vetting even before investing the sizeable observing program for the RVs themselves.
Happily, the TFOP has a wide variety of researchers, infrastructure, and resources to put toward the effort.
If a candidate proves suitable for RV follow up (as was the case for TOI-2010), the mass measurement and orbital refinement come packaged together.
This provides a lot of value for the RV investment.
That being said, RV-derived periods generally have uncertainties on the scale of hours or days for single-transit targets (which tend to have longer orbits), 
and the predictive timing uncertainty on future transits only grows worse as transits go undetected.
This ephemeris is generally insufficiently refined to plan precisely timed observations (e.g., transits and eclipses), but it does open the door for the last step needed to constrain the system neatly.

The 3$\sigma$ uncertainty window of an RV-predicted transit can easily span up to 20 days or more.
Photometric instruments that are able to locate a transit within such a wide window can refine the period uncertainties to the order of minutes. 
% They just need the rough RV-derived period to define this window.
The space-based vantage point provided by NEOSSat and other small space telescopes \citep[e.g., CHEOPS;][]{CHEOPS} is ideal for this application.
Such facilities can monitor this star over the whole time frame, and their detections do not even have to be of particularly high SNR. 
The single high-quality TESS measurement is generally sufficient to constrain the transit shape, 
and so a low SNR additional transit can simply supply timing information beyond the precision of the RVs.
Being able to point at a target at any time 
also allows space-faring instruments to quickly narrow down possible period aliases if two widely spaced transits have been detected.

It is a major and ongoing challenge to carry out successful retransit searches for TESS single-transit targets.
The nature of the mission's sector-by-sector and hemisphere-by-hemisphere observing strategy leaves plenty of room for longer-period planets to fall through the cracks.
Small space telescopes are uniquely suited in providing support observations to pull these long-period planets back from the edge of obscurity by firmly establishing their ephemerides.

\subsection{Cause of the RV Acceleration}

We believe that the 1\farcs9 Gaia neighbour is unlikely to be the cause of the residual RV slope for two reasons.
% \cmc{(should I look deeper into Gaia stuff -- e.g. chat with Clémence?)}
%
Firstly, the Gaia DR3 parallax distances of TOI-2010 and this faint neighbour differ by 1.5\,pc.  
This does place the source as a close neighbour in interstellar terms, but given their respective parallax uncertainties their distances are more than $2\sigma$ discrepant. 
It is therefore very likely to be physically separated from TOI-2010 at the parsec scale and thus could not cause the observed acceleration given its low magnitude-inferred mass.
Secondly, even if we assume some bias on the parallax measure and that they are indeed at the same distance as one another, this companion does not fall in a favourable location of the Figure~\ref{fig:planet2} plot.  The source's inferred mass and orbital separation place it 
away from the high-probability region.
%
% pixel value: 0.053
% peak at this mass: 0.787
%
The specific relative probability of its location is only $\sim$5\% of the global peak probability, and $\sim$7\% of the highest-probability region at its particular mass.
The combination of these two reasons disfavour the Gaia star as the source of the acceleration.

The 1\farcs5 source found in the `Alopeke image
may be a slightly better candidate, though it is missing some crucial information.  With a similar inferred mass but tighter separation, this source is closer to the high-likelihood region of the Figure~\ref{fig:planet2} plot. 
Its particular cell is $\sim$12\% of the peak probability for this mass. The uncertainty on its semi-major axis allows for it to intersect a bit deeper into the high-probability region.  
However, with an unknown parallax, it is entirely possible that this is a background or foreground object and wholly unassociated with the system.

Based on their low-probability locations in the search grid, coupled with a parsec-scale difference in distance between the Gaia neighbour and TOI-2010 and lack of parallax information on the `Alopeke neighbour, it appears unlikely that either star is responsible for the RV acceleration.

The presence of these two sources certainly does not rule out the additional possibility of a hidden lower-mass object.  
The `Alopeke source may be approaching the right region of parameter space, but the unknown nature of its 3D location relative to TOI-2010 precludes any certainty for now.  
Further characterization of the nearby sources may offer more clarity, such as refinement of the Gaia star's parallax, or a check on the `Alopeke star's proper motion in a few years time.
Additional RV coverage of TOI-2010 may even reveal some clear curvature to the residual acceleration, which would add strong constraints on the high-mass/wide-orbit end of the currently allowed parameter space.

%% --------------------------------------------------- %%
\section{Summary}\label{Sec:Summary}

In the course of this study, we have confirmed the planetary nature of the exoplanet TOI-2010\,b.  
A wide range of data sets from the TESS mission and the TFOP working group were collected in this effort.  
Most notably, the initial single transit discovered in the TESS Sector 15 data provided strong transit morphology constraints, but no information on the period.  
A substantial RV campaign involving several observatories mapped the RV curve, determined a rough period, and predicted a subsequent transit to an uncertainty of a few days.  
Using NEOSSat, we observed a continuous week-long window and caught this transit, refining the period down to just a few minutes uncertainty.  
A fortuitous catch in TESS's Sector 56 light curve revealed an additional transit detection at a late stage of this manuscript's preparation.

We carried out a global model fit using {\tt EXOFASTv2} to determine the system's physical and orbital parameters by simultaneously fitting time-series light curves, RVs, and historical photometric data. 
TOI-2010\,b turns out to be Jupiter-like in size, about 30\% more massive, and its equilibrium temperature may fluctuate between roughly 360--450 K given its moderately eccentric orbit.  

Our bulk metallicity analysis also suggests a fairly Jupiter-like metal mass fraction, i.e. modestly lower than the general trend given its mass \citep{Thorngren_etal_2016}.
The host star is very slightly super-solar in terms of mass, radius, luminosity, and temperature.

%As for a takeaway:  this is a fairly typical cool Jupiter, with a metallicity modestly below the expected amount for its mass (roughly Z = .165, Thorngren 2016).  In that regard it is similar with Jupiter, which similarly has a somewhat low metallicity compared to the exoplanet population.  The main differences between the planets are the younger age of TOI-2010\,b, and the higher incident flux, which both indicate that Helium rain will not have begun on TOI-2010\,b as it likely has on Jupiter.

We find evidence of a small-amplitude residual acceleration in the RV data set once TOI-2010\,b's signal has been removed, potentially indicative of an outer companion in the system.
Searching a broad grid of potential mass and semi-major axis values, we determine the relative probability that such companions could cause the observed acceleration. 
We also determine which of these simulated systems would be visible in our high-contrast imaging.
Among hidden objects (too faint and/or close to TOI-2010 to be detected), we find a correlated allowed parameter space ranging 0.4--475 $M_{\rm J}$ in mass (475\,$M_{\rm J} \approx 0.45 M_\odot$) and 2.6--23\,au in semi-major axis along its highest-probability region.  
Smaller masses and orbits cannot reproduce the observed RV slope, while more massive (i.e. brighter) objects on wider orbits would be observable in our Gemini-N/`Alopeke speckle imaging.

%
% Low end:            2.6 au  0.44 M_J
% High end visible: 
%        left -> 14.9 au   575 M_J (0.55 Msun)
%         mid -> 22.5 au   478 M_J (0.46 Msun)
%       right -> 74.7 au   275 M_J (0.26 Msun)
%
%

We make note of two nearby sources that could potentially be connected to the acceleration.  
A Gaia source, 1\farcs9 away, is at a similar distance to the TOI-2010 system, but perhaps not close enough to be considered a binary capable of producing the RV slope.  Also, its inferred mass and semi-major axis do not fall in a likely region of the parameter space.  A second source, discovered in our high-contrast imaging 1\farcs5 away, has an inferred mass and semi-major axis that are slightly more likely to produce the acceleration.  However, with no parallax information, this source could easily be just a projected neighbour.

In refining TOI-2010\,b's period, we have enabled future transit/eclipse-based research.
In large part due to its cool temperature, the transmission spectroscopy potential of this target is somewhat poor.  However, its predicted signal strength is much better for emission spectroscopy and RM measurements.

%%% Brief summary of planet's context

% (This could be moved to discussion, leaving only a small mention of final results).
TOI-2010\,b turns out to be very typical member of the population of known Jovian exoplanets.  
However, the planet's unique value and interest stem from its observability and low insolation/effective temperature.  
Currently, amongst giant planets with reliable radius and mass measurements, only $\sim$20
are at comparable or lower stellar insolation levels. 
Among those, only two are bright targets.

% Final sign-off statemtents ??
TOI-2010\,b is a successful case of searching for and catching additional transits for a single-transit candidate planet.  This process is often expensive and challenging, but it allows us to build up the confirmed exoplanet catalogue where it is only sparsely populated.

% PROBLEMS reading in "median.tex"

% Table is too wide because there are more columns
% for extra instruments (CoRoT is really NEOSSat)
%   (particularly the 3 entries in "Transit params")
%
% Have to comment out the \begin{document} and \end{document}
%
% Replace \multicolumn{3}{c}{Values}}
% with \multicolumn{3}{l}{~~~~~~~~~~~~~~~~~Values}}
% if we've got several extra columns

%% --------------------------------------------------- %%

\acknowledgements
\newpage

%% Chris and Paul 
C.R.M. and D.L. acknowledge funding from 
the Trottier Family Foundation in their support of Trottier Institute for Research on Exoplanets (iREx).  
They also acknowledge individual funding from the Natural Sciences and Engineering Research Council (NSERC) of Canada. 
P.A.D. acknowledges support by a 51 Pegasi b Postdoctoral Fellowship from the Heising-Simons Foundation and by a National Science Foundation (NSF) Astronomy and Astrophysics Postdoctoral Fellowship under award AST-1903811. 

%% Shweta Dalal
S.D. is funded by the UK Science and Technology Facilities Council (grant No. ST/V004735/1).

%% Xavier Delfosse and Thierry Forveille
X.D. and T.Fo. acknowledge funding from the French National Research Agency in the framework of the Investissements d Avenir program (ANR-15-IDEX-02), through the funding of the ``Origin of Life" project of the Grenoble-Alpes University.

%% Eder Martioli
E.M. acknowledges funding from FAPEMIG under project number APQ-02493-22 and research productivity grant No. 309829/2022-4 awarded by the CNPq, Brazil.

%% Diana Dragomir
D.D. acknowledges support from the NASA Exoplanet Research Program grant 18-2XRP18\_2-0136, and from the TESS Guest Investigator Program grants 
80NSSC22K1353 and 80NSSC22K0185.

%% Tara Fetherolf
T.Fe. acknowledges support from the University of California President's Postdoctoral Fellowship Program.

%% Karen Collins and Dave Latham
K.A.C. and D.W.L. acknowledge support from the TESS mission via subaward s3449 from MIT.

%% GMU observers that didn't make author cut
The authors would like to thank the on-duty telescope observers Patrick Newman, Owen Alfaro, Ben Chang, and William McLaughlin for their contribution in gathering the George Mason University Observatory data.

%% TESS data
This paper made use of data collected by the TESS mission and are publicly available from the Mikulski Archive for Space Telescopes (MAST) operated by the Space Telescope Science Institute (STScI). Funding for the TESS mission is provided by NASA’s Science Mission Directorate.

%% TSO and SPOC  
We acknowledge the use of public TESS data from pipelines at the TESS Science Office and at the TESS Science Processing Operations Center.

Resources supporting this work were provided by the NASA High-End Computing (HEC) Program through the NASA Advanced Supercomputing (NAS) Division at Ames Research Center for the production of the SPOC data products.

%% GI proposals that put target on 2-min/20-sec cadence lists
We would like to thank the PIs of the TESS Guest Investigator programs that put TOI-2010 on the 2 minute 
(Steven Villanueva -- G04195, 
Diana Dragomir -- G04231, 
Andrej Prsa -- G04171, 
Andrew Mayo -- G04242, and
James Davenport -- G04039) and 20\,s
(Guadalupe Tovar Mendoza -- G05121, and  Daniel Huber -- G05144) 
cadence lists.

%% ExoFOP
This research has made use of the Exoplanet follow up Observation Program 
(ExoFOP; doi:\dataset[10.26134/ExoFOP5]{https://exofop.ipac.caltech.edu/tess/})
website, which is operated by the California Institute of Technology, under contract with the National Aeronautics and Space Administration under the Exoplanet Exploration Program.

%% TSTPC
We would like to thank and acknowledge the efforts of the TESS Single Transit Planet Candidate working group for working to keep tabs on and improving our understanding of long-period targets.

%% LCOGT
This work makes use of observations from the LCOGT network. Part of the LCOGT telescope time was granted by NOIRLab through the Mid-Scale Innovations Program (MSIP). MSIP is funded by the NSF.

%% Gemini-Alopeke
Observations in the paper (Program ID: GN-2021A-LP-105) made use of the high-Resolution imaging instrument `Alopeke. `Alopeke was funded by the NASA Exoplanet Exploration Program and built at the NASA Ames Research Center by Steve B. Howell, Nic Scott, Elliott P. Horch, and Emmett Quigley. `Alopeke was mounted on the Gemini-North telescope of the international Gemini Observatory, a program of NSF’s NOIRLab, which is managed by the Association of Universities for Research in Astronomy  (AURA) under a cooperative agreement with the National Science Foundation on behalf of the Gemini Observatory partnership: the National Science Foundation (United States), National Research Council (Canada), Agencia Nacional de Investigaci\'{o}n y Desarrollo (Chile), Ministerio de Ciencia, Tecnolog\'{i}a e Innovaci\'{o}n (Argentina), Minist\'{e}rio da Ci\^{e}ncia, Tecnologia, Inova\c{c}\~{o}es e Comunica\c{c}\~{o}es (Brazil), and Korea Astronomy and Space Science Institute (Republic of Korea).

%% SOPHIE
This work is based on observations collected with the SOPHIE spectrograph on the 1.93\,m telescope at the Observatoire de Haute-Provence (CNRS), France. We thank the staff of the Observatoire de Haute-Provence for their support at the 1.93\,m telescope and on SOPHIE.
% (from Guillaume)
We also thankfully acknowledge grants from CNES and the CNRS ``Programme National de Planétologie."

%% NASA Exoplanet Archive
This research has made use of the NASA Exoplanet Archive, which is operated by the California Institute of Technology, under contract with the National Aeronautics and Space Administration under the Exoplanet Exploration Program.

%% ESA
This work has made use of data from the European Space Agency  Gaia (\url{https://www.cosmos.esa.int/gaia}), processed Data Processing and Analysis Consortium (DPAC; \url{https://www.cosmos.esa.int/web/gaia/dpac/consortium}). Funding for the DPAC has been provided by national institutions, in particular the institutions participating in the Gaia Multilateral Agreement.

\software{
{\tt lightkurve} \citep{Lightkurve_2018},
{\tt EXOFASTv2} \citep{Eastman:2019}, 
{\tt ReaMatch} \citep{ReaMatch2015}, 
{\tt astropy} \citep{astropy:2013,astropy:2018,astropy:2022}, 
{\tt matplotlib} \citep{Hunter_2007}, 
{\tt Numpy} \citep{numpy_2011,numpy_2020}
{\tt tpfplotter}\footnote{\url{https://github.com/jlillo/tpfplotter}},
{\tt ExoFile}\footnote{\url{https://github.com/AntoineDarveau/exofile}},
{\tt KeplerSpline} \citep{Vanderburg_etal_2016},
{\tt TESS-SIP} \citep{Hedges_etal_2020},
{\tt SpeckMatch} \citep{SpecMatch_Synthetic_thesis,SpecMatch_2017,SpecMatch_Empirical},
{\tt AstroImageJ} \citep{AIJ_2017},
{\tt alnitak}\footnote{ \url{https://github.com/oalfaro2/alnitak}},
{\tt AUSTRAL} \citep{Endl_etal_2000},
{\tt BANZAI-NRES} \citep{BANZAI-NRES},
{\tt PyAstronomy}\footnote{\url{https://github.com/sczesla/PyAstronomy}} \citep{PyAstronomy}.
}

\newpage
\appendix{

The RV data used in the global fit are displayed in Table~\ref{tab:rvs}.  The data span three instruments and include their individual RV offsets.

    %
% CHRIS: we can leave all lines or only the first 10 and include the table note. Up to you.
%
\startlongtable
\begin{deluxetable*}{cccc}
\tabletypesize{\scriptsize}
\tablecaption{RV Measurements of TOI-2010 \label{tab:rvs}}
\tablehead{
  \colhead{BJD$_{\rm TDB}$} & 
  \colhead{RV (m s$^{-1}$)} &
  \colhead{$\sigma_{\rm RV}$ (m s$^{-1}$)} &
  \colhead{Inst.}}
  \startdata
2458887.037868 & $-$65.7 & 8.1 & Levy \\
2458895.027687 & $-$71.6 & 6.5 & Levy \\
2458899.012701 & $-$35.9 & 5.6 & Levy \\
2458922.023014 & $-$8.4 & 6.4 & Levy \\
2458955.028682 & 39.1 & 4.4 & Levy \\
2458961.908021 & 41.6 & 6.6 & Levy \\
2458964.893612 & 48.0 & 5.0 & Levy \\
2458966.898471 & 31.0 & 9.0 & Levy \\
2458970.887634 & 38.8 & 6.5 & Levy \\
2458973.892349 & 47.6 & 5.1 & Levy \\
2458991.906717 & 10.7 & 8.9 & Levy \\
2459030.819021 & $-$42.9 & 5.4 & Levy \\
2459040.997378 & $-$22.5 & 4.5 & Levy \\
2459064.767218 & $-$6.9 & 4.7 & Levy \\
2459068.806213 & $-$0.4 & 4.4 & Levy \\
2459182.597717 & $-$38.5 & 5.6 & Levy \\
2459185.604636 & $-$26.5 & 5.0 & Levy \\
2459188.603420 & $-$34.3 & 4.6 & Levy \\
2459194.590484 & $-$25.0 & 6.9 & Levy \\
2459203.597023 & $-$12.1 & 5.9 & Levy \\
2459222.590712 & 0.7 & 5.9 & Levy \\
2459225.067097 & 46.2 & 8.8 & Levy \\
2459228.601064 & 30.8 & 7.6 & Levy \\
2459235.060004 & 25.8 & 6.1 & Levy \\
2459262.996513 & 45.1 & 9.2 & Levy \\
2459274.930264 & 22.9 & 6.7 & Levy \\
2459287.907247 & $-$33.7 & 6.3 & Levy \\
2459295.883380 & $-$40.4 & 6.7 & Levy \\
2459301.842307 & $-$47.0 & 6.3 & Levy \\
2459307.004259 & $-$55.4 & 4.3 & Levy \\
2459315.878742 & $-$28.7 & 4.7 & Levy \\
2459319.949170 & $-$36.2 & 4.0 & Levy \\
2459322.978069 & $-$34.2 & 4.4 & Levy \\
2459326.986152 & $-$33.7 & 5.4 & Levy \\
2459333.953995 & $-$27.9 & 4.7 & Levy \\
2459336.970295 & $-$14.1 & 5.7 & Levy \\
2459339.999687 & $-$20.1 & 4.5 & Levy \\
2459344.872338 & 9.5 & 5.1 & Levy \\
2459347.947959 & 0.7 & 4.3 & Levy \\
2459351.983896 & $-$3.0 & 5.6 & Levy \\
2459356.940402 & 19.2 & 4.8 & Levy \\
2459361.893688 & 22.2 & 4.4 & Levy \\
2459364.922348 & 23.0 & 4.9 & Levy \\
2459368.925800 & 9.8 & 5.0 & Levy \\
2459372.940835 & 46.4 & 7.7 & Levy \\
2459381.765521 & 26.3 & 5.5 & Levy \\
2459385.822790 & 45.6 & 4.4 & Levy \\
2459388.885495 & 46.5 & 6.5 & Levy \\
2459392.788945 & 35.9 & 5.0 & Levy \\
2459395.795554 & 51.9 & 4.3 & Levy \\
2459398.811073 & 60.4 & 4.5 & Levy \\
2459401.841438 & 37.1 & 4.6 & Levy \\
2459404.841232 & 40.2 & 4.8 & Levy \\
2459407.851764 & 23.1 & 4.1 & Levy \\
2459409.840484 & 26.1 & 4.3 & Levy \\
2459410.773285 & 24.0 & 4.4 & Levy \\
2459412.737338 & 28.5 & 4.4 & Levy \\
2459415.750915 & 16.4 & 4.8 & Levy \\
2459419.751557 & $-$11.5 & 4.6 & Levy \\
2459424.700659 & $-$8.0 & 5.0 & Levy \\
2459427.744671 & $-$19.6 & 4.7 & Levy \\
2459434.772695 & $-$23.3 & 4.4 & Levy \\
2459438.778047 & $-$59.5 & 4.6 & Levy \\
2459452.747840 & $-$47.4 & 4.5 & Levy \\
2459464.768780 & $-$32.5 & 5.1 & Levy \\
2459464.768780 & $-$32.5 & 5.1 & Levy \\
2459489.826089 & $-$9.1 & 4.8 & Levy \\
2459522.781338 & 44.9 & 6.1 & Levy \\
2459586.597180 & $-$35.6 & 5.6 & Levy \\
2459607.020932 & $-$38.8 & 6.4 & Levy \\
2459039.4716 & $-$15360.0 & 3.0 & SOPHIE \\
2459113.4148 & $-$15277.0 & 3.0 & SOPHIE \\
2459133.4087 & $-$15305.0 & 6.0 & SOPHIE \\
2459394.5368 & $-$15278.0 & 4.0 & SOPHIE \\
2459419.5450 & $-$15304.0 & 5.0 & SOPHIE \\
2459445.4901 & $-$15378.0 & 3.0 & SOPHIE \\
2459469.4911 & $-$15340.0 & 3.0 & SOPHIE \\
2459477.4045 & $-$15323.0 & 7.0 & SOPHIE \\
2459483.4480 & $-$15324.0 & 5.0 & SOPHIE \\
2459501.3575 & $-$15312.0 & 3.0 & SOPHIE \\
2459548.2221 & $-$15276.0 & 5.0 & SOPHIE \\
2459553.2376 & $-$15298.0 & 4.0 & SOPHIE \\
2459559.2231 & $-$15299.0 & 3.0 & SOPHIE \\
2459567.2305 & $-$15330.0 & 3.0 & SOPHIE \\
2459571.2387 & $-$15348.0 & 3.0 & SOPHIE \\
2459631.7057 & $-$15312.0 & 3.0 & SOPHIE \\
2459662.6759 & $-$15259.0 & 4.0 & SOPHIE \\
2459731.5828 & $-$15359.0 & 3.0 & SOPHIE \\
2459750.5120 & $-$15352.0 & 3.0 & SOPHIE \\
2459773.5751 & $-$15290.0 & 3.0 & SOPHIE \\
2459786.6144 & $-$15290.0 & 4.0 & SOPHIE \\
2459805.5426 & $-$15267.0 & 9.0 & SOPHIE \\
2459811.3339 & $-$15241.0 & 4.0 & SOPHIE \\
2459828.3791 & $-$15264.0 & 3.0 & SOPHIE \\
2459191.579631 & 8750.7 & 12.1 & Tull \\
2459192.588106 & 8747.0 & 10.8 & Tull \\
2459275.950854 & 8817.7 & 13.3 & Tull \\
2459276.970138 & 8798.1 & 10.6 & Tull \\
2459277.969833 & 8777.6 & 10.0 & Tull \\
2459293.930022 & 8740.0 & 10.4 & Tull \\
2459294.957330 & 8737.9 & 11.5 & Tull \\
2459301.941670 & 8729.9 & 13.7 & Tull \\
2459339.884675 & 8763.0 & 11.0 & Tull \\
2459340.824938 & 8757.6 & 10.4 & Tull \\
2459355.874441 & 8794.3 & 11.6 & Tull \\
2459383.844963 & 8823.0 & 11.2 & Tull \\
2459384.820734 & 8811.8 & 10.0 & Tull \\
2459411.737865 & 8795.2 & 9.6 & Tull \\
2459413.773637 & 8798.0 & 9.0 & Tull \\
2459878.627818 & 8753.2 & 11.6 & Tull \\
\enddata
%\tablenotetext{}{This is a representative subset of the full data set. The full table will be made available in machine readable format.}
\end{deluxetable*}
}

%% --------------------------------------------------- %%
% \bibliographystyle{aasjournal}
\bibliography{main.bib}

\end{document}